\documentclass[superscriptaddress,onecolumn,secnumarabic,nobibnotes,aps,prd,showpacs,nofootinbib,12pt]{revtex4}
\usepackage{eurosym}
\usepackage{graphicx}
\usepackage{epsf}
\usepackage{bm}
\usepackage{amsmath}
\usepackage{amsfonts}
\usepackage{amssymb}
\usepackage{color}
\usepackage{subfig}

\providecommand{\U}[1]{\protect\rule{.1in}{.1in}}

\newcommand{\be}{\begin{equation}}
\newcommand{\ee}{\end{equation}}

\newcommand{\mincir}{\raise
-3.truept\hbox{\rlap{\hbox{$\sim$}}\raise4.truept\hbox{$<$}\ }}
\newcommand{\magcir}{\raise
-3.truept\hbox{\rlap{\hbox{$\sim$}}\raise4.truept\hbox{$>$}\ }}

\begin{document}

\title{Compartmentalization and Coexistence in the Dark Sector of the
Universe}
\author{Andronikos Paliathanasis}
\email{anpaliat@phys.uoa.gr}
\affiliation{Institute of Systems Science, Durban University of Technology, Durban 4000,
South Africa}
\affiliation{Centre for Space Research, North-West University, Potchefstroom 2520, South
Africa}
\affiliation{Departamento de Matem\'{a}ticas, Universidad Cat\'{o}lica del Norte, Avda.
Angamos 0610, Casilla 1280 Antofagasta, Chile}
\affiliation{School of Technology, Woxsen University, Hyderabad 502345, Telangana, India}
\author{Kevin Duffy}
\email{kevind@dut.ac.za}
\affiliation{Institute of Systems Science, Durban University of Technology, Durban 4000,
South Africa}
\author{Amlan Halder}
\email{amlanhalder1@gmail.com}
\affiliation{School of Technology, Woxsen University, Hyderabad 502345, Telangana, India}
\author{Amare Abebe}
\affiliation{Centre for Space Research, North-West University, Potchefstroom 2520, South
Africa}
\affiliation{National Institute for Theoretical and Computational Sciences (NITheCS),
South Africa}

\begin{abstract}
We revise the cosmological interaction between dark energy and dark matter.
More precisely, we focus on models that support compartmentalization or
coexistence in the dark sector of the universe. Within the framework of a
homogeneous and isotropic, spatially flat
Friedmann--Lema\^itre--Robertson--Walker geometry, we analyse the asymptotic
behaviour of the physical parameters for two interacting models, where dark
energy and dark matter have constant equations of state parameters, in the
presence of dark radiation, when dark energy is described by a quintessence
scalar field. For each model, we determine the asymptotic solutions and
attempt to understand how the interaction affects the cosmological evolution
and history.
\end{abstract}

\maketitle
\date{\today }

%\title{something good title?}

\section{Introduction}

\label{intro}

The nature of dark energy and dark matter is a mystery. These exotic fluids
do not interact with light, and we observe only the gravitational effects
they are responsible for; they form the so-called dark sector of the
universe. Dark matter has been introduced to explain the velocity profile of
galaxies \cite{dm1}. On the other hand, dark energy \cite{de1} has been
introduced to explain cosmic acceleration \cite{rr1,Kowal}. In contrast to
dark matter, which exerts gravitational effects, dark energy exerts
repulsive forces that lead to anti-gravitational effects. In gravitational
physics, matter sources are described by the formalism of fluid dynamics.

On large scales, where the universe is isotropic and homogeneous, as
described by the Friedmann--Lema\^itre--Robertson--Walker (FLRW) geometry,
dark matter is described by a pressureless fluid source, also known as dust,
while dark energy is described by a perfect fluid with a negative equation
of state parameter \cite{de1}.

The cosmological constant provides the simplest mechanism for describing
dark energy \cite{rr1}. Despite the fact that the cosmological constant
introduces the fewest degrees of freedom and is in agreement with the
majority of cosmological data, it suffers from various problems. For a
recent discussion on the challenges of the cosmological constant, we refer
the reader to \cite{lpe1}. In order to overcome the puzzles of the
cosmological constant, cosmologists have dealt with the introduction of
other dark energy models with a dynamical equation for the state parameter.
The origin of the new dynamical variables is mainly attributed to the
existence of scalar fields \cite{nik,gal02,hor,in01,in02,in03,in06,in07}, or
to modifications of the gravitational action integral \cite%
{mod0,mod00,mod7,Ferraro,Ferraro2,in08,on2,Clifton,revh}.

Interaction in the dark sector of the universe \cite{Amendola:1999er}, that
is, energy transfer between dark matter and dark energy, has been proposed
as a potential mechanism to explain the cosmic coincidence problem \cite%
{con1,con2,con3,con4}. Indeed, the existence of this interaction leads to
new dynamical behaviour of the physical variables. Because of an interaction
it is possible the nature of the dark energy fluid can change. Although the
interaction between dark energy and dark matter is mainly a phenomenological
consideration, there are various theoretical models that predict an
interaction, such as a Weyl integrable spacetime \cite{salim,ww01}, the
Chameleon mechanism \cite{ch1,am1}, a conformal transformation \cite{sf3,sf4}%
, Chiral scalar fields \cite{qq2,ancqg,atr6,vr91} and others. The analysis
of the cosmological observations with the interacting models has shown that
the existence of a nonzero energy transfer term between these two fluids can
be a potential solutions for the cosmological tensions, that is, the $H_{0}$
tensions and the $\sigma _{8}$ tension \cite%
{Kumar:2017dnp,DiValentino:2017iww,span,Kumar:2019wfs,Pourtsidou:2016ico,An:2017crg,luca1,poi}%
. There is a plethora of interacting models which have been proposed in the
literature, see for instance \cite%
{ss1,ss2,ss3,ss4,ss5,ss6,ss7,ss8,ss9,ss10,ss11,ss12,ss14,ss15,ss16,qq,ss17}
and references therein.

In biological systems, understanding dynamic interactions and predicting
future states can be quite complex due to the multitude of interacting
components and variables. Compartmental models serve as a powerful tool in
simplifying these complexities by segregating a biological system into
interconnected, homogeneous compartments representing different states or
classes of elements within the system. Compartmental models can provide
detailed quantitative analyses by defining transfer rates between
compartments, which is crucial for understanding system dynamics and
predicting future states, especially useful in epidemiology. These models
also allow for robust parameter estimation and can be adapted or generalized
for various studies. Additionally, they integrate data from multiple
sources, enhancing analysis accuracy and robustness across fields like
ecology and epidemiology. A similar approach is used in this study, where
dark matter and dark energy are considered as compartments of variable
composition. An additional term in the energy conservation equations
represents the energy exchange between these compartments. Analyses of the
differential equations of the system are used to explore the long-term
behavior of the universe, using techniques to simplify the equations.

In this study, we employ dynamical system analysis to investigate the
evolution of the physical parameters for some dark energy and dark matter
interacting models which describe compartmentalization or coexistence in the
dark sector of the universe. The analysis of phase space is a powerful
method for studying the asymptotic behaviour of a given cosmological model,
as well as for the reconstruction of the provided cosmological history \cite%
{dyn1,dyn2,dyn3,dyn4,dyn5,dyn6,dyn7,dyn8,dyn9,dyn10}. The method has been
widely applied in the analysis of background equations in model cosmology,
as well as in the analysis of the asymptotic behaviour of cosmological
perturbations \cite{dp1,dp2}. The structure of the paper is as follows.

In Section \ref{sec1}, we introduce the cosmological model under our
consideration, which is that of an isotropic and homogeneous spatially flat
FLRW geometry. Moreover, the continuity equation for the cosmological fluid
is discussed, and the concept of interaction is introduced. In Section \ref%
{sec2}, we assume that the cosmological fluid consists of dark energy and
dark matter. These two fluids are assumed to have constant equation-of-state
parameters, but they have a nonzero interaction term. We consider two
interacting functions which can describe compartmentalization or coexistence
between the two fluids in the dark sector of the universe. We employ the $H$%
-normalization approach and perform a detailed analysis of the solution
space for the field equations. In particular, we calculate the stationary
points and determine their stability properties, as well as the physical
properties of the asymptotic solutions.

In order to make our analysis more practical, in Section \ref{sec3}, we
introduce a radiation term. The existence of radiation leads to a new
behaviour in the evolution of the dynamics. Moreover, in Section \ref{sec4},
we replace the minimally coupled radiation fluid with that of dark
radiation. This is an exotic matter source which interacts with the other
two fluids of the dark sector of the universe. In Section \ref{sec5}, we
assume that the dark energy fluid is described by a scalar field minimally
coupled to gravity. In this framework, dark energy has a dynamical equation
of state parameter. We focus only on the case where dark matter exists and
investigate the solution space. Because of the new dynamical variables
introduced by the scalar field, the evolution of the physical solutions
differs from those found before. Finally, in Section \ref{sec6}, we
summarize the results and draw our conclusions.

\section{The Cosmological model}

\label{sec1}

According to the cosmological principle, the geometry which describes the
universe on large scales is the spatially flat FLRW metric with line element
\begin{equation}
ds^{2}=-dt^{2}+a^{2}(t)\left( dx^{2}+dy^{2}+dz^{2}\right)\;.  \label{metric1}
\end{equation}%
Function $a(t)$ describes the radius of a three-dimensional space, and the
expansion is given by the Hubble function $H=\frac{\dot{a}}{a}$,~$\dot{a}=%
\frac{da}{dt}$. The line element (\ref{metric1}) is isotropic and
homogeneous where from the Einstein tensor%
\begin{equation}
G_{\mu \nu }\equiv R_{\mu \nu }-\frac{R}{2}g_{\mu \nu }\;,  \label{g.01}
\end{equation}
only the diagonal terms survive%
\begin{equation}
G_{\mu \nu }=diag\left( 3H^{2},-2\dot{H}-3H^{2},-2\dot{H}-3H^{2},-2\dot{H}%
-3H^{2}\right)\;.  \label{g.02}
\end{equation}
The gravitational field equations of Einstein's General Relativity are
\begin{equation}
G_{\mu \nu }=T_{\mu \nu },  \label{g.03}
\end{equation}%
where $T_{\mu \nu }$ is the energy momentum tensor which attributes the
matter source in the universe.

For the FLRW geometry, and for the Einstein tensor (\ref{g.02}) it is
obvious that $T_{\mu \nu }$ inherits the properties of $G_{\mu \nu }$;
consequently, the cosmological fluid is describe by an isotropic and
homogeneous perfect fluid, i.e.%
\begin{equation}
T_{\mu \nu }=\left( \rho +p\right) u_{\mu }u_{\nu }+pg_{\mu \nu }\;,
\label{g.04}
\end{equation}%
where $u^{\mu }$ is the comoving observer, $u^{\mu }=\delta _{t}^{\mu }$, $%
\rho \left( t\right) $ is the energy density and $p\left( t\right) $ is the
pressure component. Hence, the field equations are%
\begin{eqnarray}
3H^{2} &=&\rho \;,  \label{g.05} \\
2\dot{H}+3H^{2} &=&-p\;.  \label{g.06}
\end{eqnarray}%
The Bianchi identity gives $G_{~~~;\nu }^{\mu \nu }=0$, from where we
calculate the equation of motion for the cosmological fluid~%
\begin{equation}
T_{~~;\nu }^{\mu \nu }=0\;,  \label{g.0a}
\end{equation}%
or for the FLRW geometry%
\begin{equation}
\dot{\rho}+3H\left( \rho +p\right) =0\;.  \label{g.07}
\end{equation}%
The effective equation of state parameter for the cosmological fluid is
defined as $w_{eff}=\frac{p}{\rho }$. With the application of the field
equations the latter parameter is expressed as
\begin{equation}
w_{eff}=-1-\frac{2}{3}\frac{\dot{H}}{H^{2}}.  \label{g.07a}
\end{equation}

\subsection{The Dark Sector of the Universe}

Using the latest cosmological observations, the universe is dominated by two
fluid sources known as dark matter and dark energy. These two fluids do not
absorb or radiate light and are detected only through the gravitational
phenomena related to them. In the period of galaxy formation, dark matter
dominated the universe, while today dark energy dominates.
%with an analogy of  2/3 over 1/3 respectively. \

The energy momentum tensor is defined as
\begin{equation}
T_{\mu \nu }=T_{\mu \nu }^{m}+T_{\mu \nu }^{DE}.  \label{g.08}
\end{equation}
Dark matter is the \textquotedblleft missing\textquotedblright\ matter
source in galaxy formation, whose gravitational effect can explain the
velocity profile for the stars in galaxies. At cosmological scales the
pressure component is neglected, that is,
\begin{equation}
T_{\mu \nu }^{m}=\rho _{m}u_\mu u_\nu.  \label{g.09}
\end{equation}
On the other hand, dark energy is responsible for the late-time rapid
expansion of the universe. It has a negative pressure component, which is
related to anti-gravitational forces. The energy-momentum tensor associated
with dark energy is
\begin{equation}
T_{\mu \nu }^{DE}=\left( \rho _{d}+p_{d}\right) u_{\mu}u_{\nu }+p_{d}g_{\mu
\nu },~  \label{g.10}
\end{equation}%
from where we can define the equation of state parameter $w_{d}=\frac{p_{d}}{%
\rho _{d}}$. In the following, we consider $w_{d}<0$. By replacing (\ref%
{g.08}) in (\ref{g.0a}) we find%
\begin{equation}
\left( T_{\mu \nu }^{m}g^{\nu \kappa }+T_{\mu \nu }^{DE}g^{\nu \kappa
}\right) _{;\kappa }=0\;,  \label{g.11}
\end{equation}%
which, equivalently, can be expressed as
\begin{equation}
\left( T_{\mu \nu }^{m}g^{\nu \kappa }\right) _{;\kappa }=Q~,~\left( T_{\mu
\nu }^{DE}g^{\nu \kappa }\right) _{;\kappa }=-Q.  \label{g.12}
\end{equation}%
Here, the interaction function $Q\left( t\right) $ describes the energy
transfer between the two fluids. Thus from (\ref{g.12}) in terms of
coordinates it follows%
\begin{eqnarray}
\dot{\rho}_{m}+3H\rho _{m} &=&Q, \\
\dot{\rho}_{d}+3H\left( \rho _{d}+p_{d}\right) &=&-Q.
\end{eqnarray}
When the fluids are minimally coupled to each other, $Q\left( t\right) =0$,
and from (\ref{g.12}) we find the two conservation equations%
\begin{eqnarray}
\dot{\rho}_{m}+3H\rho _{m} &=&0,  \label{g.12A} \\
\dot{\rho}_{d}+3H\left( \rho _{d}+p_{d}\right) &=&0.  \label{g.12B}
\end{eqnarray}
Nevertheless, it is possible for the two fluids to interact and the function
$Q\left( t\right) $ to be nonzero, i.e. $Q\left( t\right) \neq 0$. In this
case there exists energy transfer between the two fluids. \ For $Q\left(
t\right) >0$, there is mass transfer from dark energy to dark matter, while
for $Q\left( t\right) <0$ dark matter is converted into dark energy.

A nonzero interaction component in the dark sector of the universe is not
neglected by the cosmological data. There are studies which support the
existence of mass transfer \cite{spp1,spp2}. Although there are theoretical
models where dark energy and dark matter are two different contributions in
a universe of a generic fluid source.

\subsection{Cosmographic parameters}

There are generally two main approaches to understanding the dynamical
history of our universe. The first is a model-independent approach, which
focuses on kinematical quantities derived directly from the space-time
metric. The second approach involves fixing either the matter content of the
universe or the underlying gravitational framework to explore its dynamics.
Kinematical quantities offer greater flexibility and novelty compared to the
model-dependent approach, as they rely solely on geometrical aspects, making
the analysis more robust. The study of kinematical quantities in a
homogeneous and isotropic universe is known as cosmography \cite{Weinberg-GR}%
.

Under the background of the homogeneous and isotropic universe characterized
by the FLRW universe, one can define the kinematical quantities as \cite%
{Weinberg-GR,Visser:2003vq}

\begin{eqnarray}
&&H(t)=\frac{1}{a}\frac{da}{dt},  \notag \\
&&q(t)=-\frac{1}{a}\frac{d^{2}a}{dt^{2}}\left[ \frac{1}{a}\frac{da}{dt}%
\right] ^{-2},  \notag \\
&&j(t)=\frac{1}{a}\frac{d^{3}a}{dt^{3}}\left[ \frac{1}{a}\frac{da}{dt}\right]
^{-3},  \notag \\
&&s(t)=\frac{1}{a}\frac{d^{4}a}{dt^{4}}\left[ \frac{1}{a}\frac{da}{dt}\right]
^{-4},  \notag
\end{eqnarray}%
where $H$, $q$, $j$, $s$ are respectively called the Hubble, deceleration,
jerk, snap parameters and except for the Hubble parameter, the last three
parameters are dimensionless. With these terminology, the expansion scale
factor $a(t)$ can be written in a Taylor expansion as follows

\begin{equation*}
a(t)=a_{0}\left[ 1+H_{0}(t-t_{0})-\frac{1}{2}q_{0}^{2}(t-t_{0})^{2}+\frac{1}{%
3!}j_{0}(t-t_{0})^{3}+\frac{1}{4!}s_{0}(t-t_{0})^{4}+\mathcal{O}%
[(t-t_{0})^{5}]\right] ,
\end{equation*}%
in which any sub-index attached to any quantity refers to its present value.
Parameters $q,~j$ and $s$ in terms of the Hubble functions are defined

\begin{align}
q& =-1-\frac{\dot{H}}{H^{2}} \\
j& =\frac{\ddot{H}}{H^{3}}-3q-2 \\
s& =\frac{\dddot{H}}{H^{4}}+4j+3q\left( q+4\right) +6,
\end{align}%
equivalently, the jerk and snap parameters are expressed in terms of the
deceleration parameter as%
\begin{eqnarray}
j &=&q+2q^{2}-\frac{\dot{q}}{H}, \\
s &=&-q\left( 1+2q\right) \left( 5+3q\right) +\left( 5+6q\right) \frac{\dot{q%
}}{H}-\frac{\ddot{q}}{H^{2}}
\end{eqnarray}%
Thus, from the evolution of $q,j,s$, one can understand the expansion of the
universe (accelerating or decelerating), the rate of acceleration ($j$) and
its next derivative. Although jerk and snap parameters are asymptotic near
today in the following we derive their dynamical evolution for the
cosmological models of our consideration.

\section{Dark Matter - Dark Energy}

\label{sec2}

In this piece of study, we follow a phenomenological approach and consider
an interacting function such that the two fluids to be described by the
simplest model of \textquotedblleft competitive fluids\textquotedblright , a
Lotka-Volterra type system, or a compartmental model such as the SIR model.

\subsection{The Compartmental Interaction}

From the interacting function \cite{mag,kos}
\begin{equation}
Q_{A}\left( t\right) =Q_{0}\left( t\right) \rho _{m}\rho _{d},  \label{in.01}
\end{equation}%
introduced in \cite{qq}, equations (\ref{g.12}) become

\begin{eqnarray}
\dot{\rho}_{m}+3H\rho _{m} &=&Q_{0}\left( t\right) \rho _{m}\rho _{d},
\label{g.14} \\
\dot{\rho}_{d}+3H\left( \rho _{d}+p_{d}\right) &=&-Q_{0}\left( t\right) \rho
_{m}\rho _{d},  \label{g.15}
\end{eqnarray}%
where $Q_{0}\left( t\right) $ has dimensions of $H^{-1}$. The latter system
can be expressed as
\begin{eqnarray}
\dot{\rho}_{m}+3H\rho _{m} &=&Q_{0}\left( t\right) \rho _{m}\rho _{d},
\label{g.16} \\
\dot{\rho}_{d}+3\left( 1+w_{d}\right) H\rho _{d} &=&-Q_{0}\left( t\right)
\rho _{m}\rho _{d}.  \label{g.17}
\end{eqnarray}%
This interaction model does not suffer from early instabilities \cite{kos}.

Although the aforementioned system has the form of the Lotka-Volterra system
with non-constant coefficients, the Hubble function $H\left( t\right) $ is
related to the energy densities $\rho _{m}$, $\rho _{d}$, from the Friedmann
equation (\ref{g.05}), that is, $3H^{2}=\rho _{m}+\rho _{d}$. Consequently,
the dynamical system (\ref{g.16}), (\ref{g.17}) is a type of compartmental
model. These models are mainly applied in epidemiology and are known as SIR
models.

\subsubsection{Asymptotic analysis}

We introduce $Q_{0}\left( t\right) =\frac{\alpha }{H}$, where $\alpha $ is a
constant, and the dimensionless variables
\begin{equation}
\Omega _{m}=\frac{\rho _{m}}{3H^{2}}~,~\Omega _{DE}=\frac{\rho _{d}}{3H^{2}}%
~,~\tau =\ln a.  \label{g.18}
\end{equation}%
Recall that the dimensionless variables have the constraint $\left\{ \Omega
_{m},\Omega _{DE}\right\} \in \left[ 0,1\right] $.

Hence, equations (\ref{g.16}), (\ref{g.17}) read%
\begin{eqnarray}
\frac{d\Omega _{m}}{d\tau } &=&3\left( \alpha +w_{d}\right) \Omega
_{m}\Omega _{DE},  \label{g.19} \\
\frac{d\Omega _{DE}}{d\tau } &=&3\Omega _{DE}\left( w_{d}\left( \Omega
_{DE}-1\right) -\alpha \Omega _{m}\right) .  \label{g.20}
\end{eqnarray}%
We remark that in terms of the dimensionless variables, where the Hubble
function has been eliminated by the dynamical system, it is clear that this
model is a compartmental model.

Furthermore, from the Friedmann's equation (\ref{g.05}) it follows
\begin{equation}
\Omega _{m}+\Omega _{DE}=1.  \label{g.21}
\end{equation}%
With the use of the latter constraint we can reduce by one the dimension of
the dynamical system (\ref{g.19}), (\ref{g.20}).

Hence the final equation is%
\begin{equation}
\frac{d\Omega _{DE}}{d\tau }=3\left( \alpha +w_{d}\right) \Omega _{DE}\left(
\Omega _{DE}-1\right)\;.  \label{g.22}
\end{equation}

\paragraph{Stationary points}

We assume that $w_{d}$ is a constant, and then we calculate the stationary
points. For $\alpha +w_{d}\neq 0$, there exist two stationary points, point $%
P_{1}$ with $\Omega _{DE}\left( P_{1}\right) =0$ and $P_{2}$ with $\Omega
_{DE}\left( P_{2}\right) =1$.

The point $P_{1}$ describes an asymptotic solution where dark matter
dominates in the universe, and the effective cosmological fluid has the
equation of state parameter $w_{eff}\left( P_{1}\right) =0$. The stationary
point is an attractor for $\left( \alpha +w_{d}\right) >0$.

On the other hand, point $P_{2}$ corresponds to an asymptotic solution where
the dark energy dominates, that is, $w_{eff}\left( P_{2}\right) =w_{d}$.
Last but not least, $P_{2}$ is an attractor when $\left( \alpha
+w_{d}\right) <0$.

\subsubsection{Analytic solution}

The nonlinear differential equation (\ref{g.22}) can be solved analytically.
Indeed, the solution in terms of the scale factor $\tau =\ln a$ is given as
follows:
\begin{equation}
\Omega _{DE}\left( a\right) =\left[ 1+\Omega^{0}a^{3\left( \alpha
+w_{d}\right) }\right] ^{-1}\;,  \label{g.23}
\end{equation}
from which we construct the Hubble function%
\begin{equation}
H\left( a\right) =\hat{H}_{0}\left[ 1+\Omega^{0}a^{3\left( \alpha
+w_{d}\right) }\right] ^{\frac{w_{d}}{2\left( w_{d}+\alpha \right) }}\left[
\Omega^{0}a^{3\left( \alpha +w_{d}\right) }\left( \alpha +w_{d}\right) %
\right] ^{-\frac{1+w_{d}}{2\left( \alpha +w_{d}\right) }}\;.  \label{g.24}
\end{equation}
Here $\Omega^{0}$ and $\hat{H}_{0}$ are integration constants. %, such that
%begin{equation}
%   \Omega^0=\frac{\Omega_{m,0}}{\Omega_{DE,0}}\;,\quad
%\end{equation}
If $w_{d}=-1$, the latter analytic expression for the scale factor becomes%
\begin{equation}
H\left( a\right) =\hat{H}_{0}\left( 1+\Omega^{0}a^{3\left( \alpha -1\right)
}\right) ^{-\frac{1}{2\left( \alpha -1\right) }}.
\end{equation}

For $\alpha =0$, we recover the $\Lambda $CDM universe, that is,%
\begin{equation}
H\left( a\right) =H_{0}\left( \Omega _{\Lambda }^{0}+\Omega
_{m}^{0}a^{-3}\right) ^{\frac{1}{2}},
\end{equation}%
where we have replaced $\hat{H}_{0}=H_{0}(\Omega _{\Lambda }^{0})^{\frac{1}{2%
}}$, and $\Omega _{m}=\Omega _{\Lambda }^{0}\Omega ^{0}$.

\subsection{The Lotka-Volterra Interaction}

The Lotka-Volterra model describes coexistence. In order to recover such
behaviour between the dark energy and the dark matter we introduce the
interaction \cite{yc1}
\begin{equation}
Q_{B}\left( t\right) =Q_{0}\left( t\right) \rho _{m}\rho _{d}+Q_{1}\left(
t\right) \rho _{d},  \label{in.02}
\end{equation}%
where $Q_{0}\left( t\right) $ has the dimensions of $H^{-1}$ and $%
Q_{1}\left( t\right) $ has the dimensions of $H$.

For this interaction the continuous equations for the two fluids reads
\begin{eqnarray}
\dot{\rho}_{m}+3H\rho _{m} &=&Q_{0}\left( t\right) \rho _{m}\rho
_{d}+Q_{1}\left( t\right) \rho _{d}. \\
\dot{\rho}_{d}+3\left( 1+w_{d}\right) H\rho _{d} &=&-Q_{0}\left( t\right)
\rho _{m}\rho _{d}-Q_{1}\left( t\right) \rho _{d}.
\end{eqnarray}

We proceed with an asymptotic analysis of the field equations for this
interacting model, where we assume that $Q_{0}\left( t\right) =\frac{\alpha
}{H}$ and $Q_{1}\left( t\right) =\beta H$, with $\alpha $,$~\beta $ constant
parameters.

\subsubsection{Asymptotic analysis}

We apply the same algorithm as before and introduce dimensionless variables (%
\ref{g.18}). The field equations with dimensionless variables form the
dynamical system%
\begin{eqnarray}
\frac{d\Omega _{m}}{d\tau } &=&3\left( \alpha +w_{d}\right) \Omega
_{m}\Omega _{DE}+\beta \Omega _{DE},  \label{gg.40} \\
\frac{d\Omega _{DE}}{d\tau } &=&3\Omega _{DE}\left( w_{d}\left( \Omega
_{DE}-1\right) -\alpha \Omega _{m}\right) -\beta \Omega _{DE},  \label{gg.41}
\end{eqnarray}%
with constraint equation (\ref{g.21}).

Applying the constraint (\ref{g.21}) results in the single differential
equation%
\begin{equation}
\frac{d\Omega _{DE}}{d\tau }=\left( 3\left( \alpha +w_{d}\right) \left(
\Omega _{DE}-1\right) -\beta \right) \Omega _{DE}\text{.}  \label{gg.42}
\end{equation}

\paragraph{Stationary points}

The stationary points of equation (\ref{gg.42}) are\ the two points $\bar{P}%
_{1}$ with $\Omega _{DE}\left( \bar{P}_{1}\right) =0$ and $\bar{P}_{2}$ with
$\Omega _{DE}\left( \bar{P}_{2}\right) =1+\frac{\beta }{3\left( \alpha
+w_{d}\right) }$.

The point $\bar{P}_{1}$ describes a universe dominated by dark matter, that
is, $\Omega _{m}\left( \bar{P}_{1}\right) =1$. On the other hand, point $%
\bar{P}_{2}$ describes a universe of coexistence between the two fluids. The
solution is physically accepted when $0\leq \Omega _{DE}\left( \bar{P}%
_{2}\right)\leq1$, that is, $-1 \leq \frac{\beta }{3\left( \alpha
+w_{d}\right) } \leq0$. The equation of state parameter for the effective
fluid in solution $\bar{P}_{2}$ is derived $w_{eff}\left( \bar{P}_{2}\right)
=w_{d}\left( 1+\frac{\beta }{3\left( \alpha +w_{d}\right) }\right) $.

As far as the stability of the stationary points is concerned, point $\bar{P}%
_{1}$ is attractor when $3\left( \alpha +w_{d}\right) +\beta >0$, while $%
\bar{P}_{2}$ is an attractor when $-3\left( \alpha +w_{d}\right) <\beta \leq
0$.

We obtain similar results for the more general model of the Lotka-Volterra
family, $Q\left( t\right) =Q_{0}\left( t\right) \rho _{m}\rho
_{d}+Q_{1}\left( t\right) \rho _{m}+Q_{2}\left( t\right) \rho _{d}$.
Specifically, for this model we find two stationary points that describe
coexistence. When $Q_{1}\left( t\right) $, or $Q_{2}\left( t\right) $
becomes zero, one of the coexistence points reduced to that of $P_{1}$, or $%
P_{2}$, for the studied before for the compartmental interaction.

Furthermore, when $\alpha =0,~\beta =0$, the $\Lambda $CDM universe is
recovered for $\hat{H}_{0}=H_{0}\Omega _{\Lambda }$, and $\Omega _{m}=\Omega
_{\Lambda }\Omega ^{0}$.

\subsubsection{Analytic solution}

The analytic solution of equation (\ref{gg.42}) in terms of the scale factor
is
\begin{equation}
\Omega _{DE}\left( a\right) =\left( 1+\frac{\beta }{3\left( \alpha
+w_{d}\right) }\right) \left( 1+\Omega^{0}a^{3\left( \alpha +w_{d}\right)
+\beta }\right) ^{-1},  \label{gg.43}
\end{equation}%
and the analytic expression for the Hubble function is
\begin{equation}
H\left( a\right) =\hat{H}_{0}\left( 1+\Omega^{0}a^{3\left( \alpha
+w_{d}\right) +\beta }\right) ^{\frac{w_{d}}{2\left( \alpha +w_{d}\right) }%
}a^{-3\left( 1+w_{d}\left( 1+\frac{\beta }{3\left( \alpha +w_{d}\right) }%
\right) \right) }.  \label{gg.44}
\end{equation}

Thus, for the case where $w_{d}=-1$, the latter solution becomes%
\begin{equation}
H\left( a\right) =\hat{H}_{0}\left( 1+\Omega^{0}a^{3\left( \alpha -1\right)
+\beta }\right) ^{-\frac{1}{2\left( \alpha -1\right) }}a^{\frac{\beta }{%
\left( \alpha -1\right) }}.  \label{gg.45}
\end{equation}

The results in Eqs. \eqref{g.24} and \eqref{gg.44} show there is a large
wiggle room to address the issue of the Hubble tension (and the $\sigma _{8}$
tension). While one needs to do a proper constraint analysis on the
parameter space of such interactions using latest and cosmological upcoming
data - and this is within the scope of our upcoming investigation - it is
evident that this solution provides a richer range of possible values of $%
H_{0}$ than that provided by $\Lambda $CDM, and hence has the potential to
reduce the Hubble tension. In fact, the interaction model presented in Eqs. %
\eqref{in.01} and \eqref{in.02} can be thought of as more general cases of
the interaction model presented in \cite{dival2020} in which the authors
show such an interaction model has an observational support in reducing both
the Hubble and $\sigma _{8}$ tensions.

\section{Dark Energy - Dark Matter - Radiation}

\label{sec3}

Although the contribution of the radiation in the total cosmological fluid
can be neglected nowadays, radiation played an important role in the early
stages of the universe.\ In order to present a more realistic model, we
consider the energy tensor for the cosmological model to depend on the three
fluids,%
\begin{equation}
T_{\mu \nu }=T_{\mu \nu }^{m}+T_{\mu \nu }^{DE}+T_{\mu \nu }^{r}
\label{r.01}
\end{equation}%
where $T_{\mu \nu }^{r}$ describes the radiation component defined as
\begin{equation}
T_{\mu \nu }^{r}=\left( \rho _{r}+p_{r}\right) u_{gm}u_{\nu }+p_{r}g_{\mu
\nu }~,~p_{r}=\frac{1}{3}\rho _{r}.  \label{r.02}
\end{equation}

The conservation law (\ref{g.11}) is modified as%
\begin{equation}
\left( T_{\mu \nu }^{m}g^{\nu \kappa }+T_{\mu \nu }^{DE}g^{\nu \kappa
}+T_{\mu \nu }^{r}g^{\nu \kappa }\right) _{;\kappa }=0,  \label{r.03}
\end{equation}%
where since radiation does not interact with the rest of the dark components
of the universe, it follows%
\begin{equation}
\left( T_{\mu \nu }^{m}g^{\nu \kappa }+T_{\mu \nu }^{DE}g^{\nu \kappa
}\right) _{;\kappa }=0~,~\left( T_{\mu \nu }^{r}g^{\nu \kappa }\right)
_{;\kappa }=0,  \label{r.04}
\end{equation}%
Consequently, in a FLRW background the continuous equation for the radiation
fluid reads
\begin{equation}
\dot{\rho}_{r}+\frac{4}{3}H\rho _{r}=0.  \label{r.05}
\end{equation}

\subsection{The Compartmental Interaction}

Let us consider the compartmental interaction $Q\left( t\right) $ as given
in expression (\ref{in.01}) with $Q_{0}\left( t\right) =\frac{\alpha }{H}$.
We introduce the new dimensionless variable $\Omega _{r}=\frac{\rho _{r}}{%
3H^{2}}$ which describes the energy density for the radiation fluid, and $%
\Omega _{r}\in \left[ 0,1\right] $.

\subsubsection{Asymptotic analysis}

The cosmological field equations in terms of the dimensionless variables (%
\ref{g.18}) are%
\begin{eqnarray}
\frac{d\Omega _{m}}{d\tau } &=&3\left( \alpha +w_{d}\right) \Omega
_{m}\Omega _{DE}+\Omega _{m}\Omega _{r},  \label{r.06} \\
\frac{d\Omega _{DE}}{d\tau } &=&3\Omega _{DE}\left( w_{d}\left( \Omega
_{DE}-1\right) -\alpha \Omega _{m}\right) +\Omega _{DE}\Omega _{r},
\label{r.07}
\end{eqnarray}%
where for the first Friedmann's equation we derive the constraint
\begin{equation}
\Omega _{r}=1-\Omega _{m}-\Omega _{DE}.  \label{r.08}
\end{equation}

We proceed with the analysis of the stationary points for the
two-dimensional system (\ref{r.06}), (\ref{r.07}). For each stationary
point, we discuss the physical properties of the asymptotic solution and its
stability properties. Indeed, the equation of state parameter for the
effective fluid in terms of the dimensionless variables is
\begin{equation}
w_{eff}=w_{d}\Omega _{DE}+\frac{1}{3}\Omega _{r}\text{.}  \label{r.09}
\end{equation}

\paragraph{Stationary points}

The dynamical system (\ref{r.06}), (\ref{r.07}) admits four stationary
points $A=\left( \Omega _{m}\left( A\right) ,\Omega _{DE}\left( A\right)
\right) $, with coordinates; $A_{1}=\left( 1,0\right) $; $A_{2}=\left(
0,1\right) $; $A_{3}=\left( 0,0\right) $ and $A_{4}=\left( \frac{1-3w_{d}}{%
3\alpha },-\frac{1}{3\alpha }\right) $.

The stationary points $A_{1}$ and $A_{2}$ correspond to universes without a
radiation component, that is, $\Omega _{r}\left( A_{1,2}\right) =0$. That
is, points $A_{1}$, $A_{2}$ have the same physical properties with that of $%
P_{1}$ and $P_{2}$ found before. We linearize the dynamical system (\ref%
{r.06}), (\ref{r.07}) around the stationary points and we calculate the
eigenvalues $e_{1}\left( A_{1}\right) =-1$, $e_{2}\left( A_{1}\right)
=-3\left( w_{d}+\alpha \right) $ and $e_{1}\left( A_{2}\right) =-1+3w_{d}$, $%
e_{2}\left( A_{2}\right) =3\left( w_{d}+\alpha \right) $. Thus, the point $%
A_{1}$ is an attractor for $\alpha >0$ and $-\alpha <w_{d}<0$, while $A_{2}$
is an attractor for $\left\{ \alpha <0,w_{d}<0\right\} $, or $\left\{ \alpha
>0,w_{d}<-\alpha \right\} $.

Concerning the asymptotic solution at point $A_{3}$, it describes a universe
dominated by the radiation fluid, that is, $\Omega _{r}\left( A_{3}\right)
=0 $. The eigenvalues around the stationary point $A_{3}$ are $e_{1}\left(
A_{3}\right) =1$, $e_{2}\left( A_{3}\right) =1-3w_{d}$, which means that
point $A_{3}$ is always a source.

Finally, point $A_{4}$ describes a universe where all the fluid contributes
in the cosmological evolution. However, because we have assumed $w_{d}<0$,
it follows that the point is not physically accepted.

\begin{figure}[tbph]
\centering\includegraphics[width=1\textwidth]{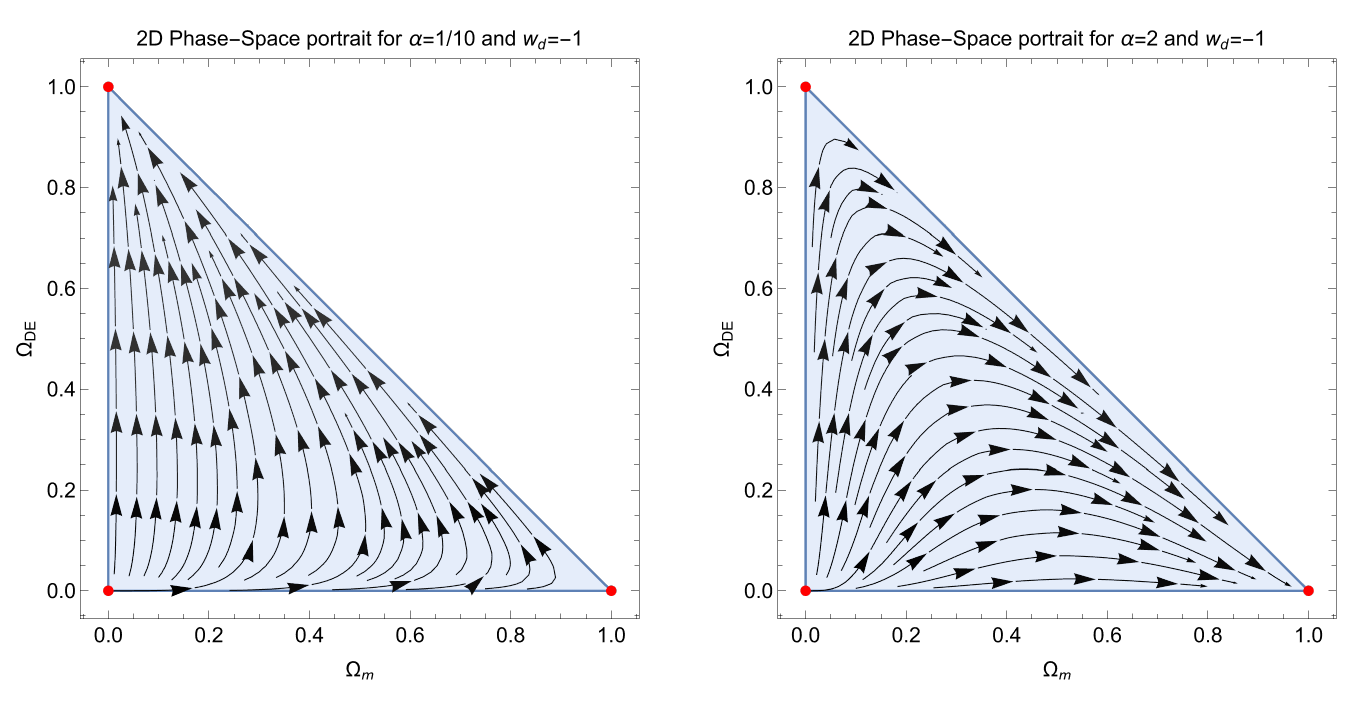}
\caption{Compartmental: Phase-space portraits for the dynamical system (%
\protect\ref{r.06}), (\protect\ref{r.07}) for $w_{d}=-1$ and $\protect\alpha %
=\frac{1}{10}$ (left fig.) and $\protect\alpha =2$ (right fig.). The
physical accepted stationary points are marked with red dots. In the left
fig. point $A_{2}$ is an attractor, while in the right fig. the attractor is
the stationary point $A_{2}$.}
\label{plot1}
\end{figure}

Two-dimensional phase-space portraits for the dynamical system (\ref{r.06}),
(\ref{r.07}) are presented in Fig. \ref{plot1}. The portraits are for $%
w_{d}=-1$ and for two values of the parameter $\alpha $, where point $A_{1}$
or $A_{2}$ are attractors. We remark that for $w_{d}=-1$ point $A_{2}$
describes the de Sitter universe.

\begin{figure}[tbph]
\centering\includegraphics[width=1\textwidth]{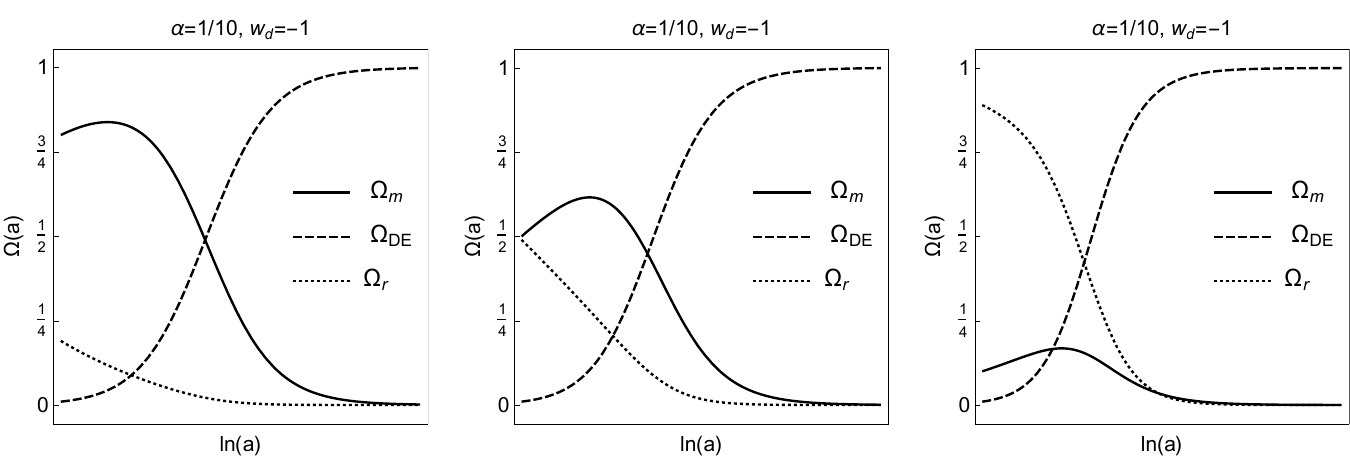}
\caption{Compartmental: Qualitative evolution of the energy density
parameters $\Omega _{m},~\Omega _{DE}$ and $\Omega _{r}$, for values of the
free parameters such that the future attractor\ of the dynamical system (%
\protect\ref{r.06}), (\protect\ref{r.07}) is the point $A_{2}$ and describe
the de Sitter solution. Plots are for different initial conditions for the
variables $\Omega _{m}\left( a\right) ,~\Omega _{DE}\left( a\right) $ and $%
\Omega _{r}\left( a\right) $. }
\label{plot2}
\end{figure}

\begin{figure}[tbph]
\centering\includegraphics[width=1\textwidth]{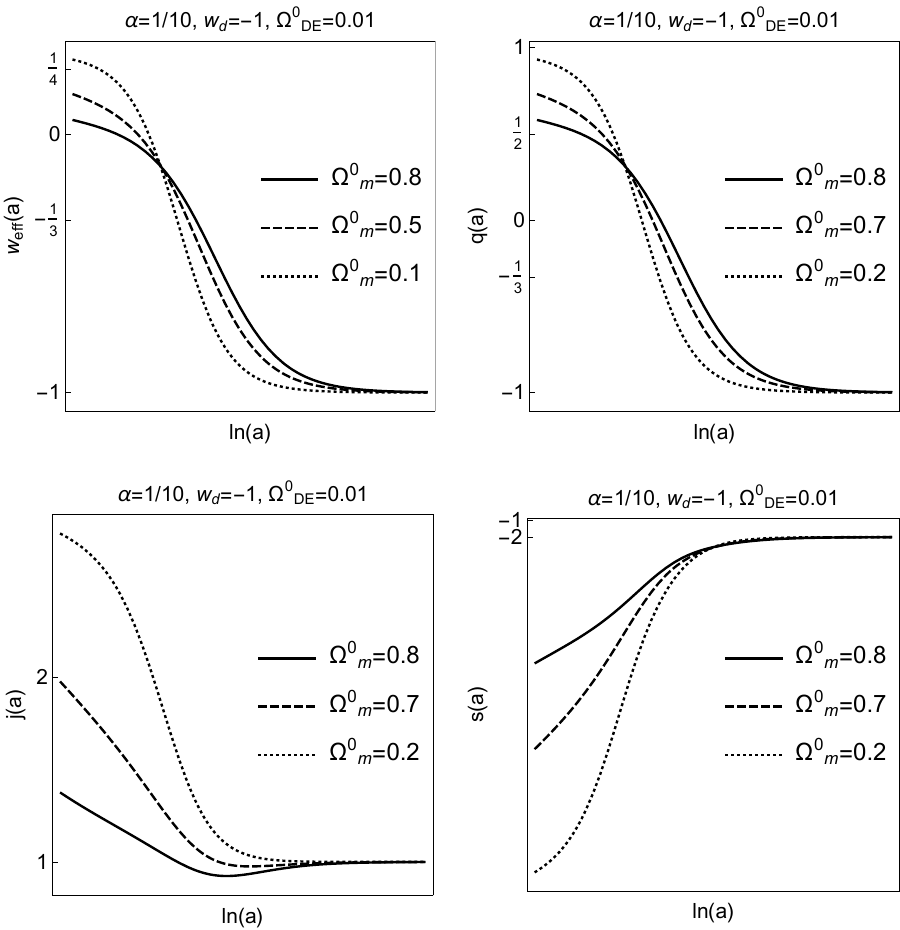}
\caption{Compartmental: Qualitative evolution of $w_{eff}$, of the
deceleration paramter $q$ and of the statefinders $j$ and $s$ for values of
the free parameters such that the future attractor for the dynamical system (%
\protect\ref{r.06}), (\protect\ref{r.07}) is the point $A_{2}$ and the
initial conditions are of that of Fig. \protect\ref{plot2}. }
\label{plot3}
\end{figure}

Furthermore, in Fig. \ref{plot2} we present the qualitative evolution for
the energy density parameters $\Omega _{m},~\Omega _{DE}$ and $\Omega _{r}$,
for values of the free parameters such that the future attractor is the
point $A_{2}$ and describes the de Sitter solution. Finally, the qualitative
evolution of the effective equation of state parameter $w_{eff}$ and of the
cosmographic parameters are given in Fig. \ref{plot3}.

\subsection{The Lotka-Volterra Interaction}

In this Section, we study the asymptotic behaviour of the field equations
for the interaction model (\ref{in.02}), which leads to a modified
Lotka-Volterra system.

\subsubsection{Asymptotic analysis}

For interaction (\ref{in.02}) the field equations in terms of the
dimensionless variables $\left\{ \Omega _{m},\Omega _{DE},\Omega _{r},\tau
\right\} $ are expressed as follows
\begin{eqnarray}
\frac{d\Omega _{m}}{d\tau } &=&3\left( \alpha +w_{d}\right) \Omega
_{m}\Omega _{DE}+\beta \Omega _{DE}+\Omega _{m}\Omega _{r},  \label{r.10} \\
\frac{d\Omega _{DE}}{d\tau } &=&3\Omega _{DE}\left( w_{d}\left( \Omega
_{DE}-1\right) -\alpha \Omega _{m}\right) -\beta \Omega _{DE}+\Omega
_{DE}\Omega _{r},  \label{r.11}
\end{eqnarray}%
with constraint equation (\ref{r.08}).

Moreover, the equation of state parameter for the effective cosmological
fluid is given by expression (\ref{r.09}). In the following section, we
investigate the stationary points of this two-dimensional system.

\paragraph{Stationary points}

The dynamical system (\ref{r.10}), (\ref{r.11}) admits the stationary points
$\bar{A}=\left( \Omega _{m}\left( \bar{A}\right) ,\Omega _{DE}\left( \bar{A}%
\right) \right) $ with coordinates $\bar{A}_{1}=\left( 1,0\right) \,\, $; $%
\bar{A}_{2}=\left( -\frac{\beta }{3\left( \alpha +w_{d}\right) },1+\frac{%
\beta }{3\left( \alpha +w_{d}\right) }\right) $; $\bar{A}_{3}=\left(
0,0\right) $ and $\bar{A}_{4}=\left( \frac{1-3w_{d}+\beta }{\alpha },-\frac{1%
}{3\alpha }+\frac{\beta }{3\alpha \left( 1-3w_{d}\right) }\right) $.

The points $\bar{A}_{1}$, $\bar{A}_{2}$ describe solutions without
radiation, i.e. $\Omega _{r}\left( \bar{A}_{1,2}\right) =0$. Hence,
asymptotic solutions have similar properties as those described by points $%
\bar{P}_{1}$ and $\bar{P}_{2}$ respectively. Furthermore, the point $\bar{A}%
_{3}$ describes a universe dominated by the radiation fluid. Finally, point $%
\bar{A}_{4}$ is physically accepted for $w_{d}=\frac{1-\beta }{3}$, with $%
\beta >1$ and $\alpha \neq 0$. However, in this case, the point $\bar{A}_{4}$
is identical to $\bar{A}_{3}$ and always describes the radiation solution.

The eigenvalues of the two-dimensional linearized system (\ref{r.10}), (\ref%
{r.11}) around the stationary points $\bar{A}_{1}$ and $\bar{A}_{2}$ are $%
e_{1}\left( \bar{A}_{1}\right) =-1$, $e_{2}\left( \bar{A}_{1}\right)
=-\left( 3\left( w_{d}+\alpha \right) +\beta \right) $; $e_{1}\left( \bar{A}%
_{2}\right) =-1+w_{d}\left( 3+\frac{\beta }{\alpha +w_{d}}\right) $, $%
e_{2}\left( \bar{A}_{2}\right) =3\left( w_{d}+\alpha \right) +\beta $.
Hence, point $\bar{A}_{1}$ is an attractor for $\alpha >-\frac{\beta }{3}$, $%
-\frac{3\alpha +\beta }{3}<w_{d}<0$; while point $\bar{A}_{2}$ is an
attractor when $a<0$ and $\beta <-3\left( \alpha +w_{d}\right) $ or $\alpha
>0$ with constraints $\left\{ w_{d}<-\alpha ,\beta <-3\left( w_{d}+\alpha
\right) \right\} $ or $\left\{ -\alpha <w_{d}<0,-\frac{\left( 1+w_{d}\right)
\left( \alpha +w_{d}\right) }{w_{d}}<\beta <-3\left( \alpha +w_{d}\right)
\right\} $.

The radiation solution described by point $\bar{A}_{3}$ is always unstable,
since the corresponding eigenvalues are calculated as $e_{1}\left( \bar{A}%
_{3}\right) =1$, $e_{2}\left( \bar{A}_{3}\right) =1-3w_{d}-\beta $. \
Finally, point $\bar{A}_{4}$ provides the eigenvalues $e_{1}\left( \bar{A}%
_{4}\right) =\frac{\alpha \beta +\sqrt{D\left( \alpha ,\beta ,w_{d}\right) }%
}{2\alpha \left( 1-3w_{d}\right) }$, $e_{2}\left( \bar{A}_{4}\right) =\frac{%
\alpha \beta -\sqrt{D\left( \alpha ,\beta ,w_{d}\right) }}{2\alpha \left(
1-3w_{d}\right) }$ with $D\left( \alpha ,\beta ,w_{d}\right) =\alpha \left(
4\left( 1-3w_{d}\right) ^{2}\left( 1-3w_{d}+\beta \right) \left( \left(
1-3w_{d}\right) \left( \alpha +w_{d}\right) -\beta w_{d}\right) +a\beta
^{2}\right) $. It follows that when the point is physically accepted, it is
always unstable.

At this point, we remark that it is possible for point $\bar{A}_{4}$ to
exist in the dynamics for other values of the free parameters $\left\{
\alpha ,w_{d},\beta \right\} $. Thus, in order to avoid nonphysical
solutions in the model parameters $w_{d}$ and $\beta $ are constrained such
that point $\bar{A}_{4}$ is identical to $\bar{A}_{3}$. However, in this
case where $w_{d}=\frac{1-\beta }{3}$, $\beta >1$, there exists a unique
attractor described by point $\bar{A}_{2}$, with $\alpha <0$. In Fig. \ref%
{plot4} we present phase-space portraits for the two-dimensional dynamical
system (\ref{r.10}), (\ref{r.11}) for $w_{d}=\frac{1-\beta }{3}$, $\beta >1$
and $\alpha <0$, where $\bar{A}_{2}$ is the unique attractor.

\begin{figure}[tbph]
\centering\includegraphics[width=1\textwidth]{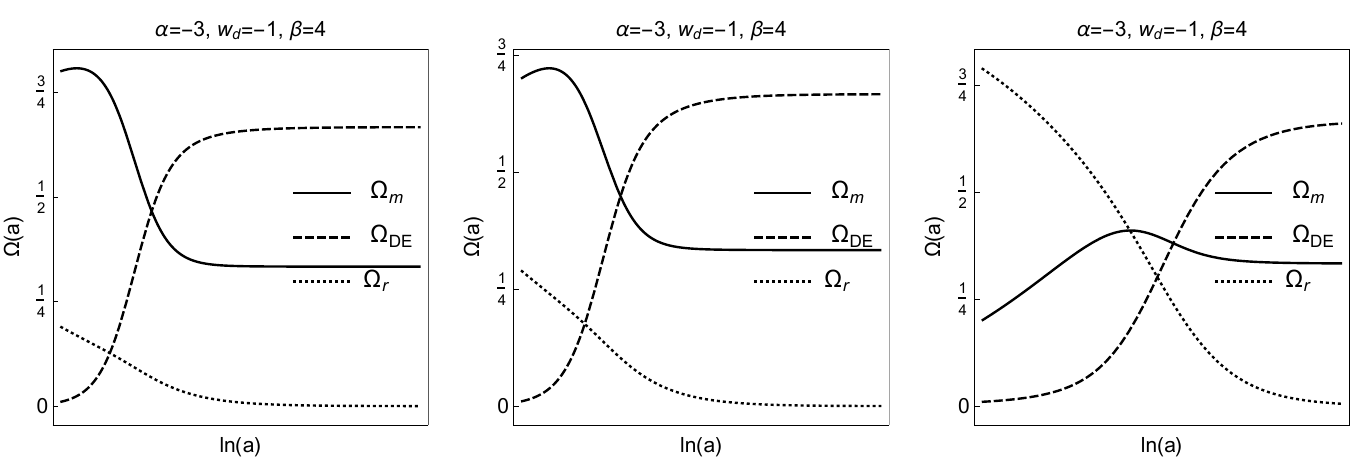}
\caption{Lotka-Volterra: Qualitative evolution of the energy density
parameters $\Omega _{m},~\Omega _{DE}$ and $\Omega _{r}$ for the dynamical
system (\protect\ref{r.10}), (\protect\ref{r.11}) for values of the free
parameters such that the future attractor is point $\bar{A}_{2}.$ \ Plots
are for different initial conditions for the variables $\Omega _{m}\left(
a\right) ,~\Omega _{DE}\left( a\right) $ and $\Omega _{r}\left( a\right) $. }
\label{plot5}
\end{figure}

\begin{figure}[tbph]
\centering\includegraphics[width=1\textwidth]{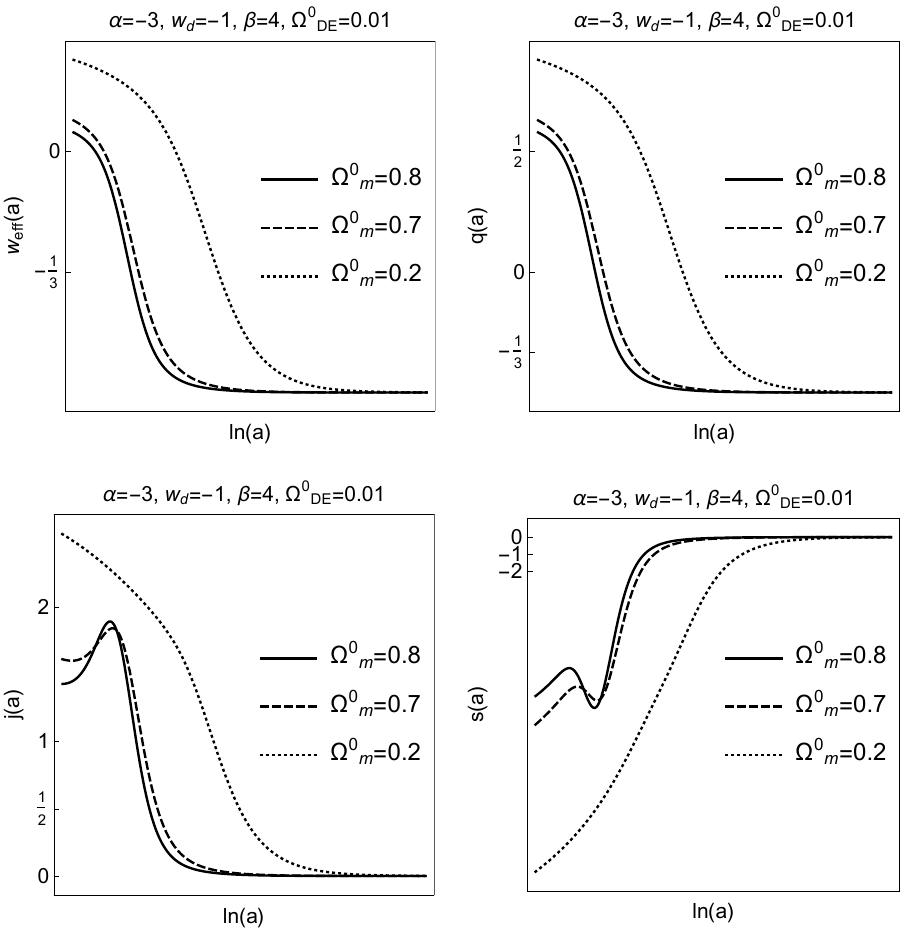}
\caption{Lotka-Volterra: Qualitative evolution of $w_{eff}$, of the
deceleration paramter $q$ and of the statefinders $j$ and $s$ for values of
the free parameters such that the future attractor of the dynamical system (%
\protect\ref{r.10}), (\protect\ref{r.11}) is point $\bar{A}_{2}$ and initial
conditions are that of Fig. \protect\ref{plot5}. }
\label{plot6}
\end{figure}

The qualitative evolution for the energy density parameters $\Omega
_{m},~\Omega _{DE}$,$~\Omega _{r}$ and of the effective equation of state
parameter $w_{eff}$ are presented in Figs. \ref{plot5} and \ref{plot6}.

Furthermore, in Fig. \ref{plot2} we present the qualitative evolution of the
energy density parameters $\Omega _{m},~\Omega _{DE}$ and $\Omega _{r}$ for
values of the free parameters such that the future attractor is the point $%
A_{2}$ and describes the de Sitter solution. Finally, the qualitative
evolution of the effective equation of state parameter $w_{eff}$ and of the
cosmographic parameters are given in Fig. \ref{plot3}.

\begin{figure}[tbph]
\centering\includegraphics[width=1\textwidth]{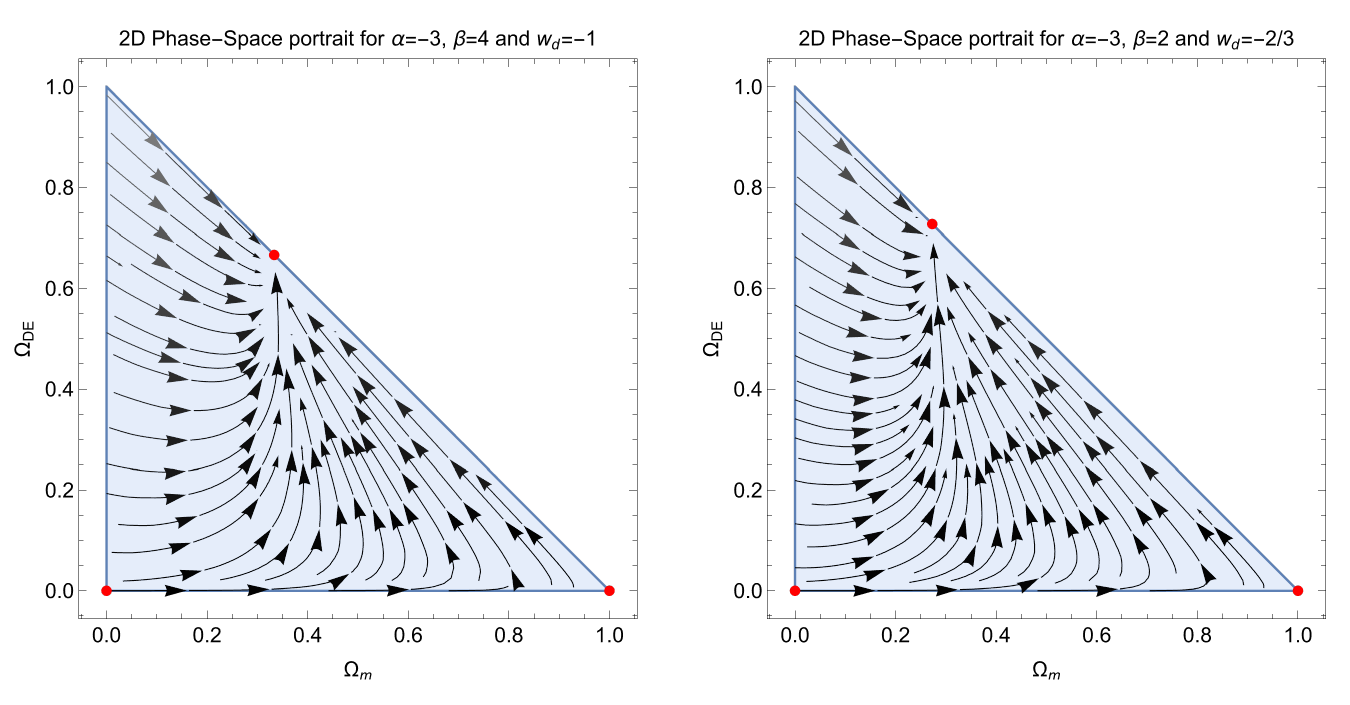}
\caption{Lotka-Volterra: Phase-space portraits for the dynamical system (%
\protect\ref{r.10}), (\protect\ref{r.11}) for $w_{d}=\frac{1-\protect\beta }{%
3}$,$~\protect\alpha =-3$ with $\protect\beta =4$ (left fig.) and $\protect%
\beta =3$ (right fig.). The physical accepted stationary points are marked
with red dots. The unique attactor in the two figures is point $\bar{A}_{2}$
which describes a universe with a coexistence of dark matter and dark
energy. }
\label{plot4}
\end{figure}

\section{Dark Energy - Dark Matter - Dark Radiation}

\label{sec4}

Dark radiation is an exotic form of radiation which has been proposed to
interact only with dark matter \cite{dr1,dr2}. Such an interaction scenario
has recently gained attention as a potential candidate for resolving the
Hubble tension \cite{dival21,gorizz23,bag24}. We will, in this section,
study the dynamical systems implications of the interaction, but we will
return to a full-fledged tensions constraining analysis vis-\`{a}-vis
cosmological data in a subsequent work.

We introduce the energy momentum tensor
\begin{equation}
T_{\mu \nu }^{R}=\left( \rho _{R}+p_{R}\right) u_{\mu}u_{\nu }+p_{R}g_{\mu
\nu }~,~p_{R}=\frac{1}{3}\rho _{R},  \label{d.01}
\end{equation}%
which we assume describes the dark radiation. The total energy momentum of
the universe is defined as
\begin{equation}
T_{\mu \nu }=T_{\mu \nu }^{m}+T_{\mu \nu }^{DE}+T_{\mu \nu }^{R},
\label{d.02}
\end{equation}%
and using the Bianchi identity we can now define the following system%
\begin{equation}
\left( T_{\mu \nu }^{m}g^{\nu \kappa }\right) _{;\kappa }=Q-V~,~\left(
T_{\mu \nu }^{DE}g^{\nu \kappa }\right) _{;\kappa }=-Q~,~\left( T_{\mu \nu
}^{R}g^{\nu \kappa }\right) _{;\kappa }=V,  \label{d.03}
\end{equation}%
where the new variable $V$ describes the interaction between dark matter and
dark radiation.

\subsection{The EMR Model}

We consider the compartmental interaction with the parameter $Q$ defined by
expression (\ref{in.01}), and we define the parameter $V$ as $V=V_{0}\left(
t\right) \rho _{m}$.

Hence, in the spatially flat FLRW background, equations (\ref{d.03}) read as%
\begin{eqnarray}
\dot{\rho}_{m}+3H\rho _{m} &=&Q_{0}\left( t\right) \rho _{m}\rho
_{d}-V_{0}\left( t\right) \rho _{m}.  \label{d.04} \\
\dot{\rho}_{d}+3\left( 1+w_{d}\right) H\rho _{d} &=&-Q_{0}\left( t\right)
\rho _{m}\rho _{d},  \label{d.05} \\
\dot{\rho}_{R}+\frac{4}{3}H\rho _{R} &=&V_{0}\left( t\right) \rho _{m}.
\label{d.06}
\end{eqnarray}

This dynamical system defines an analogue of a SIR model for cosmology. We
shall call it the dark EMR model. In an SIR model, the interaction is
between an infected population and a recovered population. In this analogue,
the interaction is between dark matter and dark radiation.

We proceed with the analysis of the asymptotics.

\subsubsection{Asymptotic analysis}

We introduce a new dimensionless variable $\Omega _{R}=\frac{\rho _{R}}{%
3H^{2}}$, and we write the cosmological field equations in an equivalent form%
\begin{eqnarray}
\frac{d\Omega _{m}}{d\tau } &=&\left( 3\left( \alpha +w_{d}\right) \Omega
_{DE}+\Omega _{R}-\gamma \right) \Omega _{m},  \label{d.07} \\
\frac{d\Omega _{DE}}{d\tau } &=&3\Omega _{DE}\left( w_{d}\left( \Omega
_{DE}-1\right) -\alpha \Omega _{m}\right) +\Omega _{DE}\Omega _{R},
\label{d.08}
\end{eqnarray}%
with constraint%
\begin{equation}
\Omega _{R}=1-\Omega _{m}-\Omega _{DE}.  \label{d.09}
\end{equation}

Finally the equation of state parameter for the total cosmological fluid is
defined as
\begin{equation}
w_{eff}=w_{d}\Omega _{DE}+\frac{1}{3}\Omega _{R}.  \label{d.10}
\end{equation}

We proceed with an analysis of the asymptotic behaviour.

\paragraph{Stationary points}

The stationary points $B=\left( \Omega _{m}\left( B\right) ,\Omega
_{DE}\left( B\right) \right) $ of the two-dimensional system (\ref{d.07}), (%
\ref{d.08}) are $B_{1}=\left( 1-\gamma ,0\right) ,~B_{2}=\left( 1,0\right) $%
, $B_{3}=\left( 0,0\right) $ and $B_{4}=\left( \frac{\left( 1-3w_{d}\right)
}{3\alpha }\left( 1-\frac{\gamma }{3\left( \alpha +w_{d}\right) }\right) ,-%
\frac{1}{3\alpha }+\frac{\gamma \left( 1+3\alpha \right) }{9\alpha \left(
\alpha +w_{d}\right) }\right) $.

The asymptotic solution at point $B_{1}$ describes a universe with dark
matter and dark radiation, i.e. $\Omega _{R}\left( B_{1}\right) =\gamma $.
The point is physically accepted for $0<\gamma <1$. This means that there is
energy transfer from dark matter to the dark radiation fluid. Points $B_{2}$
and $B_{3}$ have the same physical properties with that of points $A_{2}$
and $A_{3}$, respectively. Indeed, the point $B_{2}$ describes a universe
where only dark energy contributes to the cosmological fluid, while at the
point $B_{3}$ the asymptotic solution describes a universe dominated by the
radiation fluid.

Finally, point $B_{4}$ is an extension of the previous point $A_{4}$, where
the three fluids contribute in the universe $\Omega _{R}\left( B_{4}\right)
=1-\frac{\gamma -3w_{d}}{3\alpha }$ and $w_{eff}\left( B_{4}\right) =\frac{1%
}{3}\left( 1-\frac{\left( 1-3w_{d}\right) }{3\left( \alpha +w_{d}\right) }%
\gamma \right) $. If we require that point $B_{1}$, which describes the
matter era, exist always, that is, $0<\gamma <1$, then point $B_{4}$ is
physical accepted when $\left\{ 0<\alpha <\frac{1}{3},-\alpha <w_{d}<0,\frac{%
3\left( \alpha +w_{d}\right) }{1+3\alpha }\leq \gamma \leq 3\left( \alpha
+w_{d}\right) \right\} $, $\ $or $\alpha >\frac{1}{3}$ and$~\left\{ -\alpha
<w_{d}<\frac{\left( 1-3\alpha \right) }{3},\frac{3\left( \alpha
+w_{d}\right) }{1+3\alpha }\leq \gamma \leq 3\left( \alpha +w_{d}\right)
\right\} $ or $\left\{ \frac{\left( 1-3\alpha \right) }{3}<w_{d}<0,\frac{%
3\left( \alpha +w_{d}\right) }{1+3\alpha }\leq \gamma \leq 1\right\} $ For $%
w_{d}=-1$, the latter conditions becomes $\left\{ 1<\alpha <\frac{4}{3}~,~%
\frac{3\left( \alpha -1\right) }{1+3\alpha }\leq \gamma \leq 3\left( \alpha
-1\right) \right\} ,$ or $\left\{ \alpha >\frac{4}{3},\ ~\frac{3\left(
\alpha -1\right) }{1+3\alpha }\leq \gamma \leq 1\right\} $.

As far as the stability properties of the stationary points are concerned,
point $B_{1}$ is an attractor for $\alpha >\frac{3w_{d}-\gamma }{3\left(
\gamma -1\right) }$ while $B_{2}$ is an attractor for $\alpha <\frac{1}{3}%
\left( \gamma -3w_{d}\right) $. Furthermore, the point $B_{3}$ always
describes an unstable solution. The eigenvalues of the linearized system (%
\ref{d.07}), (\ref{d.08}) around the stationary point $B_{4}$ are $%
e_{1}\left( B_{4}\right) =\frac{\left( 3w_{d}-1\right) \alpha \gamma +\sqrt{%
\bar{D}\left( \alpha ,w_{d},\gamma \right) }}{6\alpha \left( \alpha
+w_{d}\right) }$, $e_{2}\left( B_{4}\right) =\frac{\left( 3w_{d}-1\right)
\alpha \gamma -\sqrt{\bar{D}\left( \alpha ,w_{d},\gamma \right) }}{6\alpha
\left( \alpha +w_{d}\right) }$ in which
\begin{equation}
\bar{D}\left( a,w_{b},\gamma \right) =\left( 3w_{d}-1\right) \alpha \left(
12\left( \alpha +w_{d}\right) ^{2}\left( \left( 2+3\alpha \right) \gamma
-3\left( \alpha +w_{d}\right) \right) -\alpha \left( 5+12\alpha \right)
\gamma ^{2}\right) .
\end{equation}%
When $\bar{D}\left( \alpha ,w_{d},\gamma \right) <0$, then $\frac{\left(
3w_{d}-1\right) \alpha \gamma }{6\alpha \left( \alpha +w_{d}\right) }$ and
the stationary point can be a stable sink, otherwise for $\bar{D}\left(
\alpha ,w_{d},\gamma \right) >0$ there are values of the free parameters
where the point can be an attractor. Last but not least, for the value of
the free variables where point $B_{4}$ is physical accepted, the unique
attractor is point $B_{4}$.

\begin{figure}[tbph]
\centering\includegraphics[width=1\textwidth]{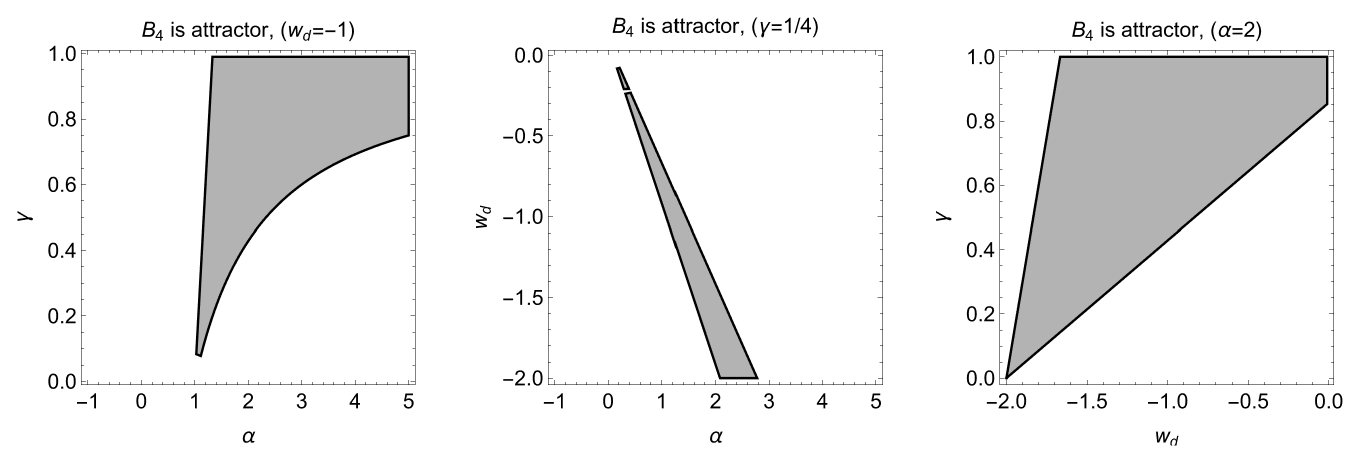}
\caption{EMR-I: Region space where the asymptotic solution described by
point $B_{4}$ is stable.}
\label{plot7}
\end{figure}

\begin{figure}[tbph]
\centering\includegraphics[width=1\textwidth]{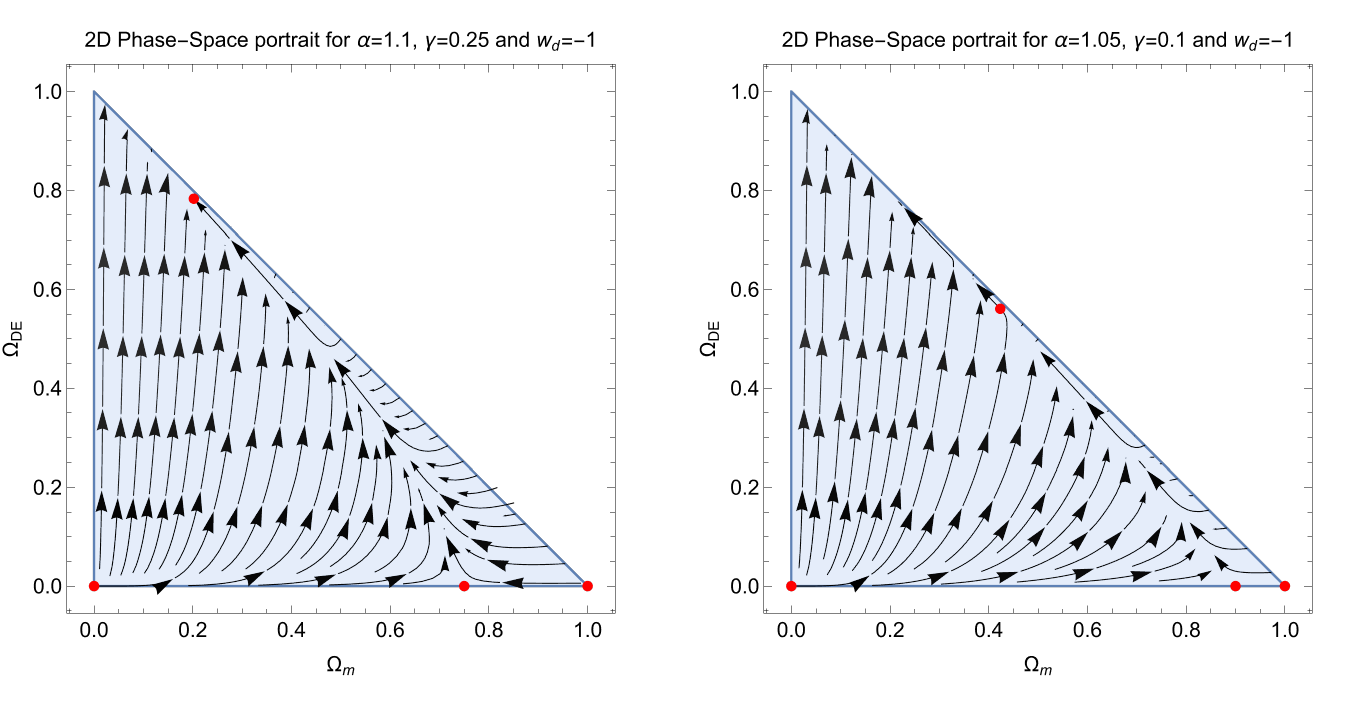}
\caption{EMR-I: Phase-space portraits for the dynamical system (\protect\ref%
{d.07}), (\protect\ref{d.08}) for values of the free parameters where $B_{4}$
is physically accepted. The dynamical space admits four stationary points
with $B_{4}$ the unique attractor, and $B_{1}$ a saddle point. }
\label{plot8}
\end{figure}

\begin{figure}[tbph]
\centering\includegraphics[width=1\textwidth]{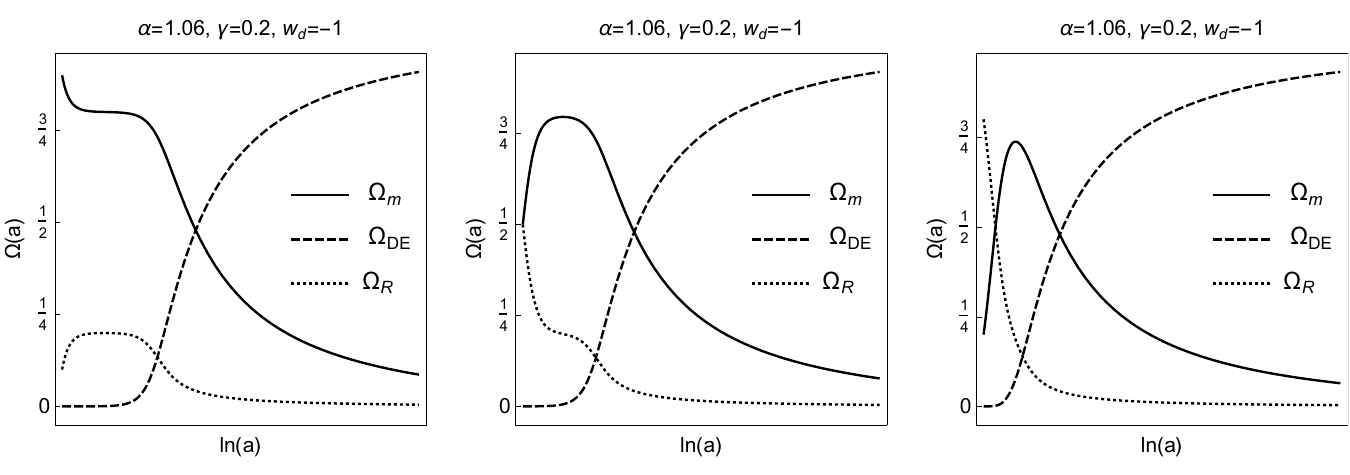}
\caption{EMR-I: Qualitative evolution of the energy density parameters $%
\Omega _{m},~\Omega _{DE}$ and $\Omega _{r}$ for the dynamical system (%
\protect\ref{d.07}), (\protect\ref{d.08}) for values of the free parameters
such that $B_{4}$ is physically accepted. Plots are for different initial
conditions for the variables $\Omega _{m}\left( a\right) ,~\Omega
_{DE}\left( a\right) $ and $\Omega _{r}\left( a\right) $. }
\label{plot9}
\end{figure}

\begin{figure}[tbph]
\centering\includegraphics[width=1\textwidth]{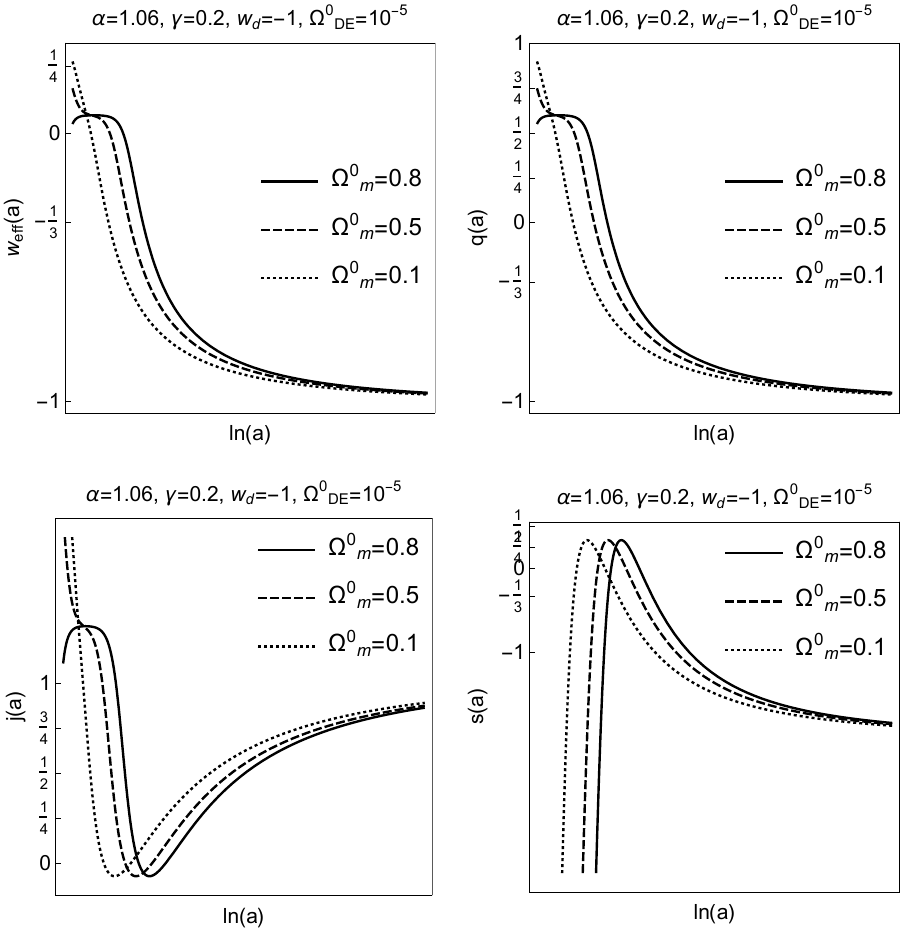}
\caption{EMR-I: Qualitative evolution $w_{eff}$, of the deceleration
paramter $q$ and of the statefinders $j$ and $s~$for values of the free
parameters such that the future attractor of the dynamical system (\protect
\ref{d.07}), (\protect\ref{d.08}) is the point $B_{4}$ and initial
conditions are that of Fig. \protect\ref{plot9}. }
\label{plot10}
\end{figure}

In Fig. \ref{plot7} we present the region space of the free parameters $%
\left( a,w_{b},\gamma \right) $, where the asymptotic solution described by
point $B_{4}$ is stable. The phase-space portrait of the dynamical system (%
\ref{d.07}), (\ref{d.08}) are given in \ref{plot8}. Moreover, the
qualitative evolution of the physical parameters $\left\{ \Omega _{m},\Omega
_{DE},\Omega _{R}\right\} $ , $w_{eff}$ and of the cosmographic parameters
are presented in \ref{plot9} and \ref{plot10} respectively.

\section{Quintessence as Dark Energy}

\label{sec5}

In the previous sections, we have assumed that the dark energy has a
constant equation of state parameter. However, this assumption suffers from
various problems. The simplest gravitational model which describes the
dynamical behaviour of the dark energy fluid is the scalar field.

Quintessence is a scalar field minimally coupled to gravity, where in the
case of homogeneous spacetimes the corresponding fluid is described by a
perfect fluid with energy density $\rho _{\phi }$ and pressure component $%
p_{\phi }$ defined as
\begin{eqnarray}
\rho _{\phi } &=&\frac{1}{2}\dot{\phi}^{2}+V\left( \phi \right) ~, \\
p_{\phi } &=&\frac{1}{2}\dot{\phi}^{2}-V\left( \phi \right) ~,
\end{eqnarray}%
where $\phi =\phi \left( t\right) $ is the scalar field and $V\left( \phi
\right) $ is the potential function which defines the scalar field mass.

The equation of state parameter for the scalar field is calculated as%
\begin{equation}
w_{\phi }=\frac{\frac{1}{2}\dot{\phi}^{2}-V\left( \phi \right) }{\frac{1}{2}%
\dot{\phi}^{2}+V\left( \phi \right) },
\end{equation}%
from where we observe that $\left\vert w_{\phi }\right\vert \leq 1$, and $%
w_{\phi }\rightarrow -1$, when the scalar field potential$~V\left( \phi
\right) $ dominates.

When there is no interaction between the scalar field and the dark matter,
equation (\ref{g.12B}) provides the equation of motion for the scalar field,
which is a second-order differential equation, that is,%
\begin{equation}
\dot{\phi}\left( \ddot{\phi}+3H\dot{\phi}+V_{,\phi }\right) =0.
\label{kg.04}
\end{equation}

In the following sections, we investigate the asymptotic dynamics for the
two models under consideration, that is, the compartmental interaction model
(\ref{in.01}) and the Lotka-Volterra interaction model (\ref{in.02}). For
simplicity of our calculations, we omit the radiation term.

\subsection{The Compartmental Interaction}

Consider the interaction in (\ref{in.01}) with $Q_{0}\left( t\right) =\alpha
\frac{\dot{\phi}}{\sqrt{6}H^{2}}$; which results in the following system%
\begin{eqnarray}
\dot{\rho}_{m}+3H\rho _{m} &=&\alpha \frac{\dot{\phi}}{\sqrt{6}}\frac{\rho
_{m}}{H^{2}}\left( \frac{1}{2}\dot{\phi}^{2}+V\left( \phi \right) \right) ,
\label{kg.05} \\
\ddot{\phi}+3H\dot{\phi}+V_{,\phi } &=&-\alpha \frac{\rho _{m}}{\sqrt{6}H^{2}%
}\left( \frac{1}{2}\dot{\phi}^{2}+V\left( \phi \right) \right) .
\label{kg.06}
\end{eqnarray}

\subsubsection{Asymptotic analysis}

We employ the $H$-normalization approach, thus, in terms of new
dimensionless variables%
\begin{equation}
x=\frac{\dot{\phi}}{\sqrt{6}H}~,~y=\frac{\sqrt{V\left( \phi \right) }}{\sqrt{%
3}H}~,~\Omega _{m}=\frac{\rho _{m}}{3H^{2}}~,~\lambda =\frac{V_{,\phi }}{V}%
~,~\tau =\ln a  \label{kg.07}
\end{equation}%
the field equations read as%
\begin{eqnarray}
\frac{dx}{d\tau } &=&\frac{1}{2}\left[ 3x\left( x^{2}-y^{2}-1\right) -\sqrt{6%
}\lambda y^{2}-3\alpha \left( x^{2}+y^{2}\right) \Omega _{m}\right] ,
\label{kg.08} \\
\frac{dy}{d\tau } &=&\frac{1}{2}y\left[ 3+\sqrt{6}\lambda x+3\left(
x^{2}-y^{2}\right) \right] ,  \label{kg.09} \\
\frac{d\lambda }{d\tau } &=&\sqrt{6}x\lambda ^{2}\left( \Gamma \left(
\lambda \right) -1\right) ~,~\Gamma \left( \lambda \right) =\frac{V_{,\phi
\phi }V}{V_{,\phi }^{2}},  \label{kg.10}
\end{eqnarray}%
with the constraint equation%
\begin{equation}
\Omega _{m}=1-x^{2}-y^{2}.  \label{kg.11}
\end{equation}%
From the constraint equation it follows that $1-x^{2}-y^{2}\geq 0$, which
means that $x,y$ take values in the two unitary disk. Furthermore, by
definition, $y>0$. Recall that the energy density for the scalar field $%
\Omega _{\phi }$ is defined as $\Omega _{\phi }=x^{2}+y^{2}$.

Last, but not least, in terms of the new variables the effective equation of
state parameter is defined as
\begin{equation}
w_{eff}\left( x,y,\lambda \right) =x^{2}-y^{2}.  \label{kg.12}
\end{equation}

For simplicity of our calculations and in order to keep the number of
dimensions and the free parameters of the dynamical system low, we assume
that the scalar field potential has a constant potential function $V\left(
\phi \right) =V_{0}$. This means that the mass of the scalar field as given
by equation (\ref{kg.06}) depends on the energy density of dark matter.

Furthermore, the parameter $\lambda $ is always zero. Thus the dynamical
evolution of the physical variables for this model is given by the
two-dimensional dynamical system
\begin{eqnarray}
\frac{dx}{d\tau } &=&\frac{1}{2}\left[ 3x\left( x^{2}-y^{2}-1\right)
-3\alpha \left( x^{2}+y^{2}\right) \left( 1-x^{2}-y^{2}\right) \right]\;,
\label{kg.14} \\
\frac{dy}{d\tau } &=&\frac{1}{2}y\left[ 3+3\left( x^{2}-y^{2}\right) \right]%
\;.  \label{kg.15}
\end{eqnarray}

\paragraph{Stationary points}

The stationary points of the dynamical system (\ref{kg.14}), (\ref{kg.15})
are defined on the two-dimensional plane $T=\left( x\left( T\right) ,y\left(
T\right) \right) $, with coordinates $T_{1}=\left( 0,0\right) $, ~$%
T_{2}^{\pm }=\left( \pm 1,0\right) $, $T_{3}=\left( -\frac{1}{\alpha }%
,0\right) $, $T_{4}=\left( 0,1\right) $ and $T_{5}=\left( x_{4}\left( \alpha
\right) ,\sqrt{1+x_{4}^{2}\left( \alpha \right) }\right) $.

The point $T_{1}$ with $\Omega _{m}\left( T_{1}\right) =1$ describes the
matter-dominated era. The linearized system around the stationary points
provides the eigenvalues $e_{1}\left( T_{1}\right) =-\frac{3}{2}$ and $%
e_{2}\left( T_{1}\right) =\frac{3}{2}$, from which we infer that the matter
solutions are always unstable and $T_{1}$ is a saddle point.

On the other hand, $T_{2}^{\pm }$\ describe solutions where the universe is
dominated by the kinetic term of the scalar field, $\Omega _{m}\left(
T_{2}^{\pm }\right) =0$, and the effective cosmological fluid is that of
stiff matter, i.e. $w_{eff}\left( T_{2}^{\pm }\right) =1$. The corresponding
eigenvalues are $e_{1}\left( T_{2}^{\pm }\right) =3$ and $e_{2}\left(
T_{2}^{\pm }\right) =3\left( 1\pm \alpha \right) $. Thus, asymptotic
solutions are always unstable. $T_{2}^{+}$ is a saddle point for $\alpha <-1$%
, otherwise it is a source, while $T_{2}^{-}$ is a saddle point for $\alpha
>1$, otherwise it is also a source.

The point $T_{3}$ describes an asymptotic solution where the kinetic term of
the scalar field and the dark matter contribute to the cosmological fluid,
i.e. $\Omega _{m}\left( T_{3}\right) =1-\frac{1}{\alpha ^{2}}$. The point is
physically accepted for $\alpha ^{2}>1$, while when $\alpha ^{2}=1$ it
reduces to the points $T_{2}^{\pm }$. For the cosmological fluid we
calculate $w_{eff}\left( T_{3}\right) =\frac{1}{\alpha ^{2}}$, which means
that it can not describe acceleration. Nevertheless, for $\alpha =\frac{1}{%
\sqrt{3}}$, point $T_{3}$ describes the radiation solution. The eigenvalues
are $e_{1}\left( T_{3}\right) =\frac{3\left( 1+\alpha ^{2}\right) }{2\alpha
^{2}}$, $e_{2}\left( T_{3}\right) =-\frac{3\left( 1-\alpha ^{2}\right) }{%
2\alpha ^{2}}$, from where we infer that the point when it is physically
accepted is a saddle point, otherwise it is a source.

The stationary point $T_{4}$ is an attractor, $e_{1}\left( T_{4}\right) =-3$%
, $e_{2}\left( T_{4}\right) =-3$, and the asymptotic solution is that of the
de Sitter universe, $w_{eff}\left( T_{4}\right) =-1$, $\Omega _{m}\left(
T_{4}\right) =0$.

Finally, for point~$T_{5},$ parameter $x_{4}$ is given by the equation $%
\alpha =\frac{1}{x_{4}\left( 1+2x_{4}^{2}\right) }$. However, $y\left(
T_{5}\right) =\sqrt{1+x_{4}^{2}\left( \alpha \right) }>1$, that is, the
point is not physically accepted. The eigenvalues of the linearized system
are $e_{1}\left( T_{5}\right) =\frac{3\left\{ -1+\sqrt{5+4x_{4}^{2}\left[
9+4x_{4}^{2}\left( 5+3x_{4}^{2}\right) \right] }\right\} }{2\left(
1+2x_{4}^{2}\right) }$, $e_{1}\left( T_{5}\right) =\frac{3\left\{ -1-\sqrt{%
5+4x_{4}^{2}\left[ 9+4x_{4}^{2}\left( 5+3x_{4}^{2}\right) \right] }\right\}
}{2\left( 1+2x_{4}^{2}\right) }$, thus $T_{5}$ is always a saddle point.

\begin{figure}[tbph]
\centering\includegraphics[width=1\textwidth]{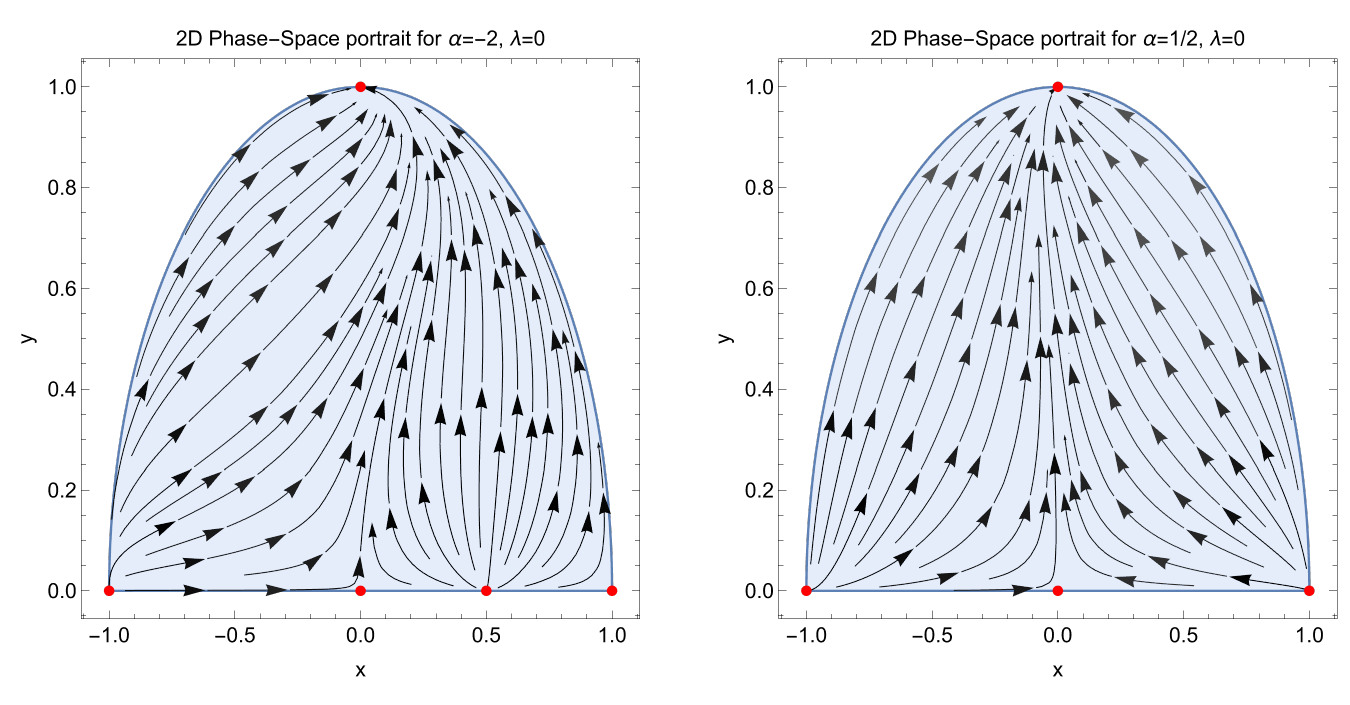}
\caption{Quintessence (Compartmental): Phase-space portraits for the
dynamical system (\protect\ref{kg.14}), (\protect\ref{kg.15}) for $\protect%
\alpha =-2$ (left fig.) and $\protect\alpha =\frac{1}{2}$ (right fig.) The
stationary points are marked with red dots. We remark that the unique
attractor is point $T_{4}$. However, the structure of the phase-space is
different when point $T_{3}$ exist. }
\label{plot11}
\end{figure}

\begin{figure}[tbph]
\centering\includegraphics[width=1\textwidth]{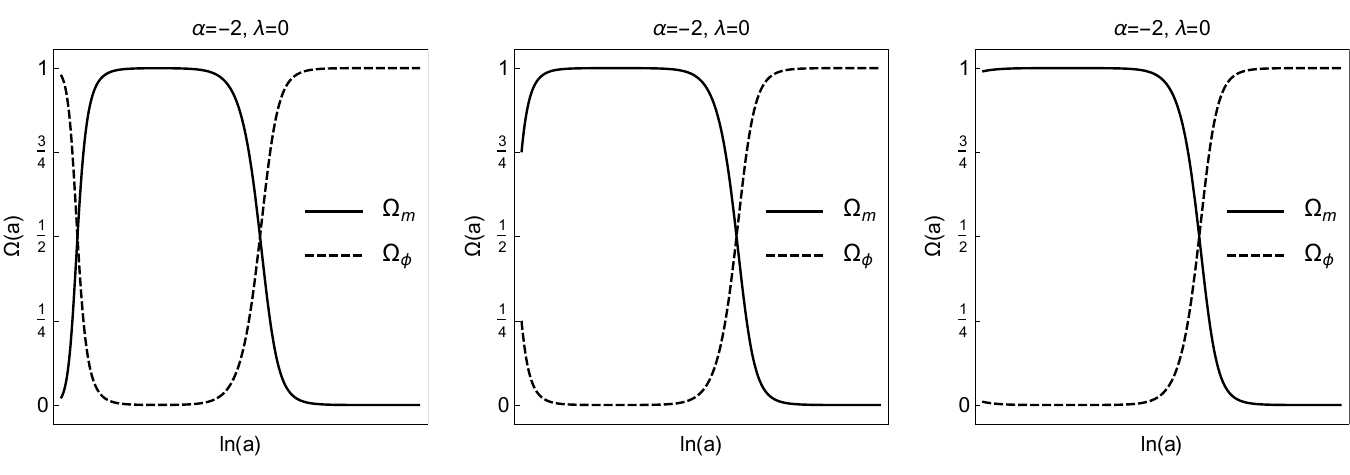}
\caption{Quintessence (Compartmental): Qualitative evolution of the energy
density parameters $\Omega _{m}$ and$~\Omega _{\protect\phi }$ the dynamical
system (\protect\ref{kg.14}), (\protect\ref{kg.15}) for $\protect\alpha =-2$%
, and different sets of initial conditions.}
\label{plot12}
\end{figure}

\begin{figure}[tbph]
\centering\includegraphics[width=1\textwidth]{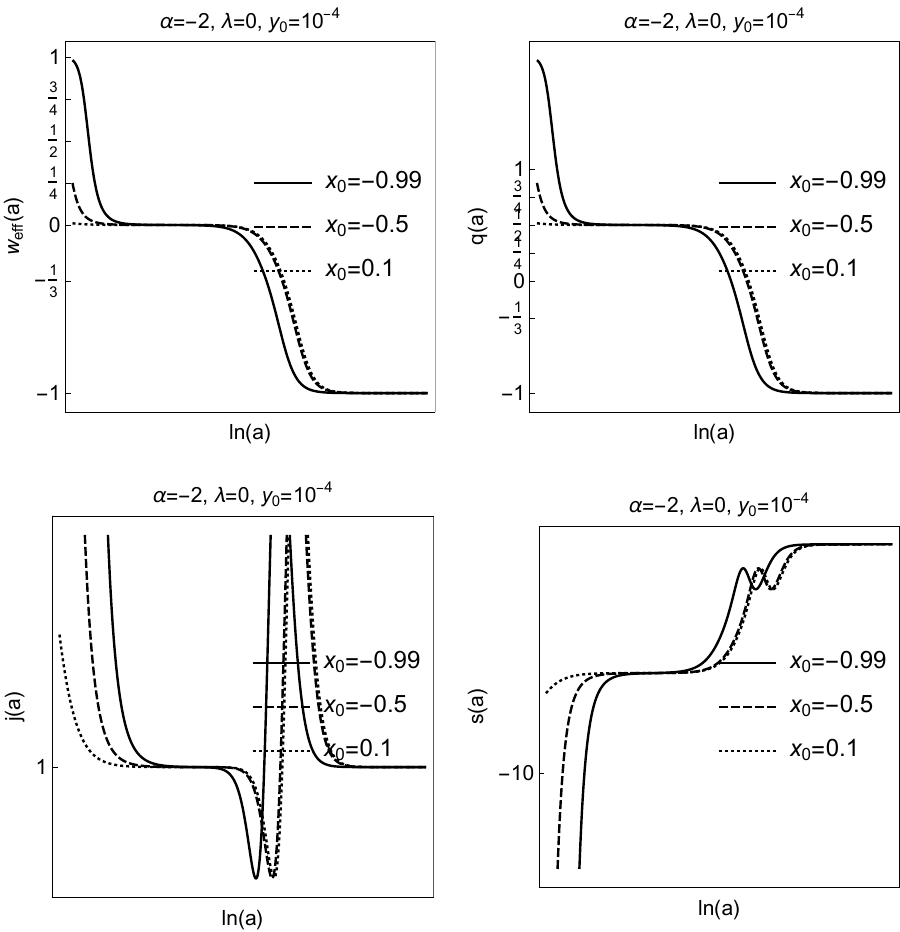}
\caption{Quintessence (Compartmental): Qualitative evolution $w_{eff}$, of
the deceleration paramter $q$ and of the statefinders $j$ and $s~~$as they
are given by the solution of the dynamical system (\protect\ref{kg.14}), (%
\protect\ref{kg.15}) and initial conditions of that of Fig. \protect\ref%
{plot12}. The de Sitter universe described by point $T_{4}$ is the unique
future attractor. }
\label{plot13}
\end{figure}

In Fig. \ref{plot11} we present phase-space diagrams for the dynamical
system (\ref{kg.14}), (\ref{kg.15}), where we observe that $T_{4}$ is the
unique attractor. Moreover, in Figs. \ref{plot12} and \ref{plot13} we
present the qualitative evolution of the physical parameters $\Omega
_{m},~\Omega _{\phi },$ $w_{eff}~$and of the cosmographic parameters for
various sets of initial conditions.

At this point, it is important to mention that for $\alpha ^{2}\leq 1$, the
physically accepted stationary points are only those of the quintessence
model without interaction. Therefore, the physically accepted stationary
points that describe non-zero interaction exist only when $\alpha ^{2}>1$.

\subsection{The Lotka-Volterra Interaction}

For the Lotka-Volterra interaction (\ref{in.02}) with $Q_{0}\left( t\right)
=\alpha \frac{\dot{\phi}}{\sqrt{6}H^{2}}$ and $Q_{1}\left( t\right) =\beta
\frac{\dot{\phi}^{2}}{H}$, in terms of the dimensionless variables the field
equations are expressed as follows%
\begin{eqnarray}
\frac{dx}{d\tau } &=&\frac{1}{2}\left[ 3x\left( x^{2}-y^{2}-1\right) -\sqrt{6%
}\lambda y^{2}-3\left( x^{2}+y^{2}\right) \left( \alpha \Omega _{m}+2\beta
x\right) \right] ,  \label{kg.16} \\
\frac{dy}{d\tau } &=&\frac{1}{2}y\left[ 3+\sqrt{6}\lambda x+3\left(
x^{2}-y^{2}\right) \right] ,  \label{kg.17} \\
\frac{d\lambda }{d\tau } &=&\sqrt{6}x\lambda ^{2}\left( \Gamma \left(
\lambda \right) -1\right) ~,~\Gamma \left( \lambda \right) =\frac{V_{,\phi
\phi }V}{V_{,\phi }^{2}}.  \label{kg.18}
\end{eqnarray}%
with algebraic constraint (\ref{kg.11}).

Hence, by applying the constraint (\ref{kg.11}) and considering a constant
potential function, that is, $\lambda =0$, we end up with the
two-dimensional system%
\begin{eqnarray}
\frac{dx}{d\tau } &=&\frac{1}{2}\left[ 3x\left( x^{2}-y^{2}-1\right) -\sqrt{6%
}\lambda y^{2}-3\left( x^{2}+y^{2}\right) \left( \alpha \left(
1-x^{2}-y^{2}\right) +2\beta x\right) \right] ,  \label{kg.19} \\
\frac{dy}{d\tau } &=&\frac{1}{2}y\left[ 3+\sqrt{6}\lambda x+3\left(
x^{2}-y^{2}\right) \right] .  \label{kg.20}
\end{eqnarray}

\paragraph{Stationary points}

The stationary points $\bar{T}=\left( x\left( \bar{T}\right) ,y\left( \bar{T}%
\right) \right) $ of the two dimensional dynamical system (\ref{kg.19}), (%
\ref{kg.20}), have the following coordinates, $\bar{T}_{1}=\left( 0,0\right)
$,~$\bar{T}_{2}=\left( 0,1\right) $, $\bar{T}_{3}=\left( x_{3}\left( \alpha
,\beta \right) ,0\right) $ and $\bar{T}_{4}=\left( x_{4}\left( \alpha ,\beta
\right) ,\sqrt{1+\left( x_{4}\left( \alpha ,\beta \right) \right) ^{2}}%
\right) $. We proceed with a discussion of the physical properties for the
asymptotic solutions at the stationary points.

The asymptotic solution at point $\bar{T}_{1}$ describes a universe
dominated by the matter era, similar to point $T_{1}$. The eigenvalues are
calculated $e_{1}\left( \bar{T}_{1}\right) =-\frac{3}{2}$, $e_{2}\left( \bar{%
T}_{2}\right) =\frac{3}{2}$, which means that $\bar{T}_{1}$ is always a
saddle point.

Moreover, point $\bar{T}_{2}$ describes the de Sitter solution as point $%
T_{4}$. The eigenvalues are $e_{1}\left( \bar{T}_{2}\right) =-3$,~$%
e_{2}\left( \bar{T}_{2}\right) =-3\left( 1+\beta \right) $. Hence, $\bar{T}%
_{2}$ is an attractor when $\beta >-1$.

For the stationary point $\bar{T}_{3}$, parameter $x_{3}\left( \alpha ,\beta
\right) $ is given by the algebraic equation $\beta =\frac{\left( 1+\alpha
x_{3}\right) \left( x_{3}^{2}-1\right) }{2x_{2}}$, where for $\beta =0$, the
points $T_{2}^{\pm }$ and $T_{3}$ are recovered. The asymptotic solution at
the point $\bar{T}_{3}$ cannot describe acceleration, i.e., $w_{eff}\left(
\bar{T}_{3}\right) =\left( x_{3}\right) ^{2}\geq 0$. We remark that $\bar{T}%
_{3}$ can describe one or three real points. We calculate the eigenvalues $%
e_{1}\left( \bar{T}_{3}\right) =\frac{3}{2}\left( 1+x_{3}^{2}\right) $ and $%
e_{2}\left( \bar{T}_{3}\right) =\frac{3}{2}\left( 2+\alpha x_{3}\left(
1+x_{3}^{2}\right) \right) $. Hence, point $\bar{T}_{3}$ always describes an
unstable solution.

Finally, the point $\bar{T}_{4}$ is not physically accepted because $y\left(
\bar{T}_{4}\right) >0$. Function $x_{4}\left( \alpha ,\beta \right) $ is
given by the algebraic equation $\beta =\alpha x_{4}-\frac{1}{1+2x_{4}^{2}}$%
. The eigenvalues of the linearized system near to the stationary point are $%
e_{1}\left( \bar{T}_{4}\right) =\frac{3}{2\left( 1+2x_{4}^{2}\right) }\left(
-1+\sqrt{1+4x_{4}\left( 1+x_{4}^{2}\right) \left( 1+2x_{4}^{2}\right) \left(
\alpha +4x_{4}\left( 1+\alpha x_{4}\left( 1+x_{4}^{2}\right) \right) \right)
}\right) $ and $e_{2}\left( \bar{T}_{4}\right) =\frac{3}{2\left(
1+2x_{4}^{2}\right) }\left( -1-\sqrt{1+4x_{4}\left( 1+x_{4}^{2}\right)
\left( 1+2x_{4}^{2}\right) \left( \alpha +4x_{4}\left( 1+\alpha x_{4}\left(
1+x_{4}^{2}\right) \right) \right) }\right) $. From these eigenvalues it is
clear that $\bar{T}_{4}$ can be an attractor. However, this will be
unphysical and the free parameters $\alpha $, $\beta $ should be constrained
such that $\bar{T}_{4}$ is a source or saddle point.

\begin{figure}[tbph]
\centering\includegraphics[width=1\textwidth]{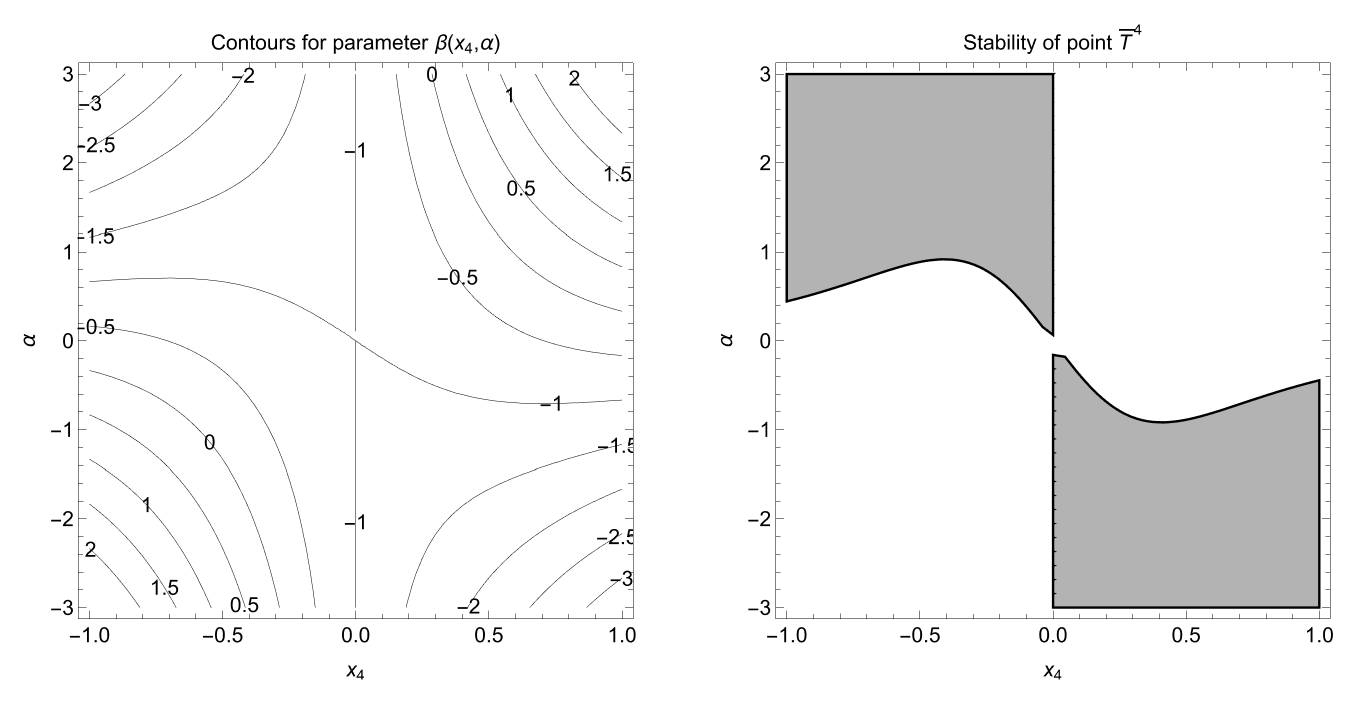}
\caption{Quintessence (Lotka-Volterra): Left fig.: Contour plot which
relates the values $x_{4},~\protect\alpha $ and $\protect\beta $. Right
fig.: Region when point $\bar{T}_{4}$ is an attractor. }
\label{plot14}
\end{figure}

\begin{figure}[tbph]
\centering\includegraphics[width=1\textwidth]{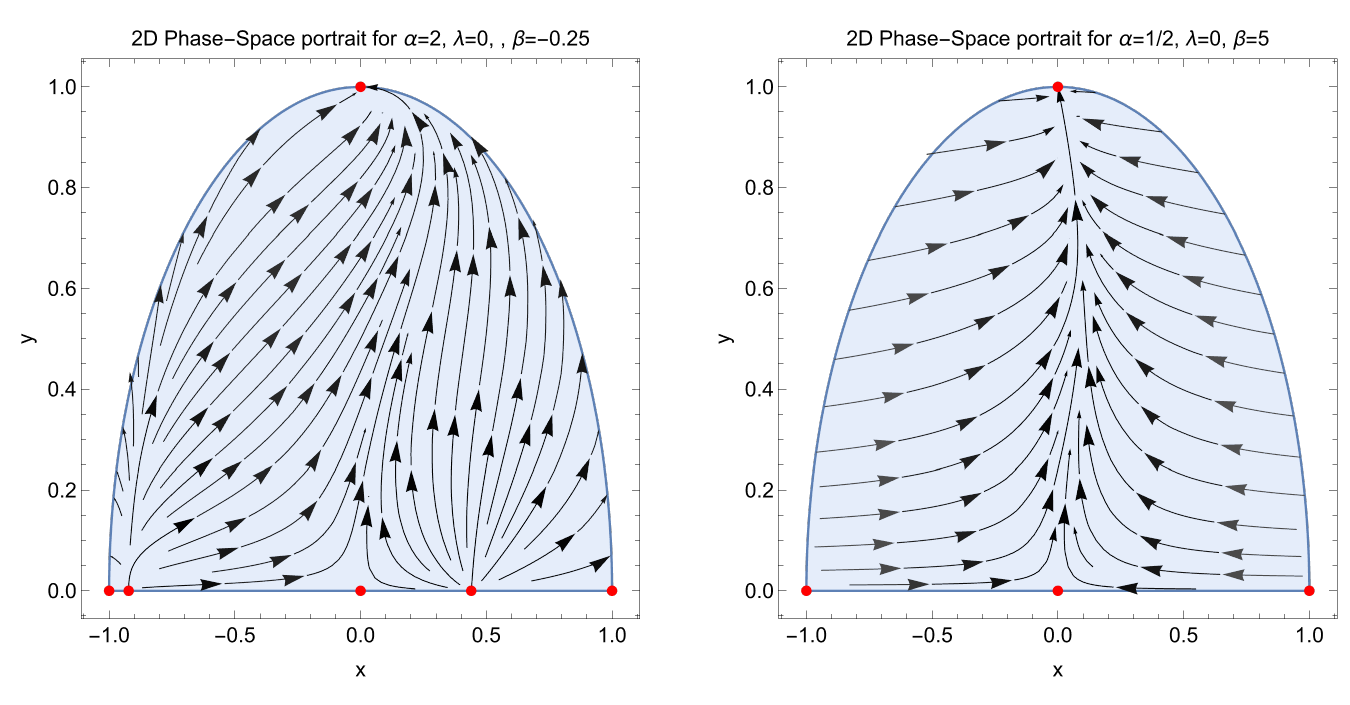}
\caption{Quintessence (Lotka-Volterra): Phase-space portraits for the
dynamical system (\protect\ref{kg.19}), (\protect\ref{kg.20}) for $\protect%
\alpha =-2$, $\protect\beta =-\frac{1}{4}$ (left fig.) and $\protect\beta =5$
(right fig.) The stationary points are marked with red dots. The unique
attractor is the de Sitter point $T_{2}$. }
\label{plot15}
\end{figure}

\begin{figure}[tbph]
\centering\includegraphics[width=1\textwidth]{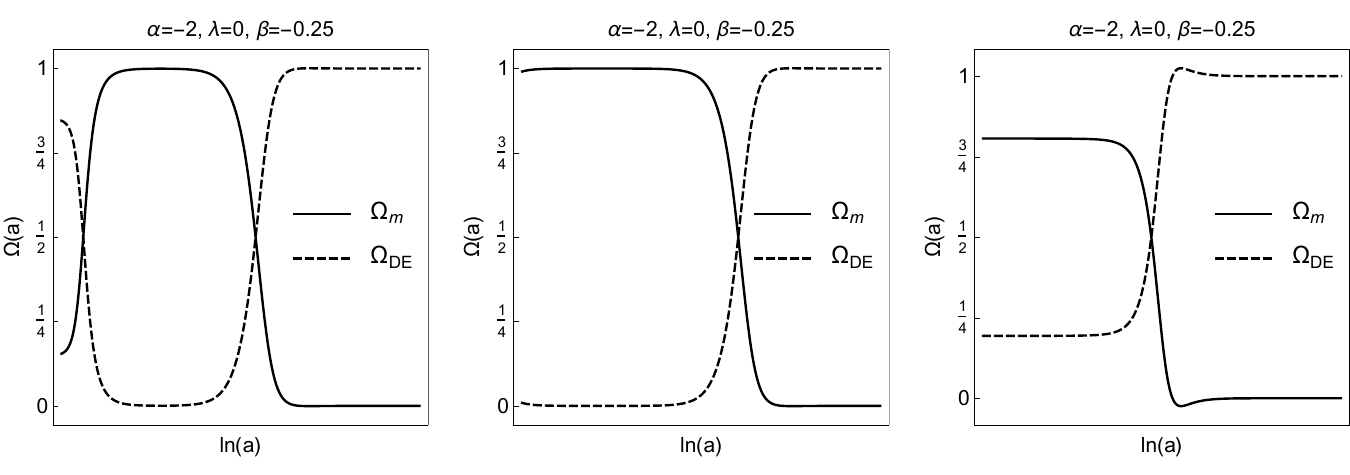}
\caption{Quintessence (Lotka-Volterral): Qualitative evolution of the energy
density parameters $\Omega _{m}$ and$~\Omega _{\protect\phi }$ the dynamical
system (\protect\ref{kg.19}), (\protect\ref{kg.20}) for $\protect\alpha =-2$%
,~$\protect\beta =-\frac{1}{4}$ and different sets of initial conditions.}
\label{plot16}
\end{figure}

\begin{figure}[tbph]
\centering\includegraphics[width=1\textwidth]{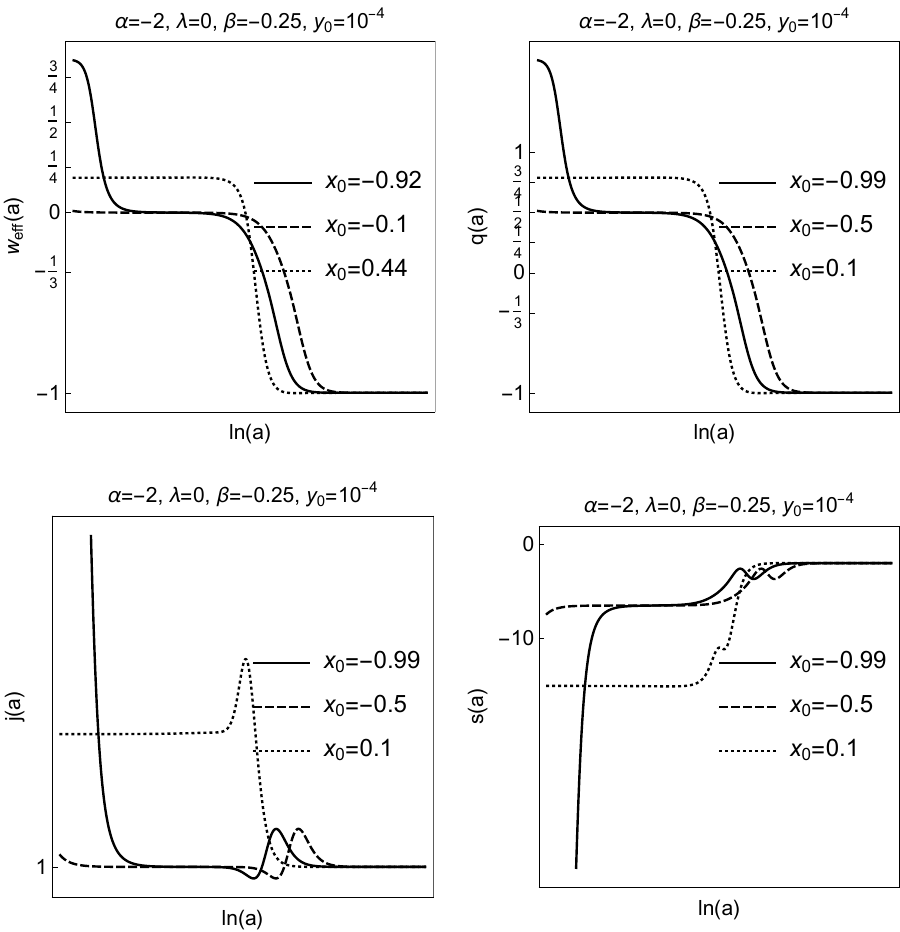}
\caption{Quintessence (Lotka-Volterra): Qualitative evolution $w_{eff}$, of
the deceleration paramter $q$ and of the statefinders $j$ and $s$ as they
are given by the solution of the dynamical system (\protect\ref{kg.14}), (%
\protect\ref{kg.15}) and initial conditions of that of Fig. \protect\ref%
{plot16}. The de Sitter universe described by point $T_{2}$ is the unique
future attractor. }
\label{plot17}
\end{figure}

In Fig. \ref{plot14} we present the contour plot that relates the free
parameters $\alpha \;,\beta $ and $x_{4}$; we also present the region where
the real parts of the eigenvalues are negative. We observe that the model is
physically accepted approximately for $\beta >-1$. Phase-space portraits of
the two-dimensional dynamical system (\ref{kg.19}), (\ref{kg.20}) are given
in Fig. \ref{plot15}. Finally, the evolution of the physical parameters $%
\Omega _{m},~\Omega _{\phi },$ $w_{eff}~$and of the cosmographic parameters
are given in Figs. \ref{plot16} and \ref{plot17}. The values of the free
parameters have been selected such that the unique attractor is the de
Sitter point $\bar{T}_{2}$.

\section{Conclusions}

\label{sec6}

Cosmological models with interaction in the dark sector are crucial for
describing the universe. Due to the interaction, new degrees of freedom are
introduced into the gravitational model, which drive the dynamics and help
explain the observations.

In this study, we examined gravitational models that describe
compartmentalization and coexistence in the dark sector of the universe.
Drawing inspiration from population dynamics models in biology, we
analogized dark energy and dark matter as two species evolving in the
``universe." Furthermore, dark radiation was introduced as the third
``species."

To study the interaction models, we employed dynamical analysis and a
phase-space investigation. We used the H-normalization approach and
calculated the asymptotic solutions and their stability properties in terms
of dimensionless variables. Each asymptotic solution corresponds to a
specific stationary point of the cosmological field equations, representing
a distinct era of cosmic evolution.

We considered dark matter to be described by a pressureless fluid source,
while for dark energy we assumed two cases. In the first case, dark energy
is described by an ideal gas with a constant equation of state parameter. In
the second scenario, we considered dark energy as described by a
quintessence scalar field. Due to the potential function, the dynamics of
the second cosmic scenario leads to a more complex cosmological history.
From this analysis, we can provide constraints on the dynamical variables of
the interacting model and discuss the cosmological viability of the models.

From this research, the interactions between dark matter and dark energy
could significantly influence the dynamics of the universe. The study
demonstrates that these interactions could potentially address cosmological
tensions such as $H_{0}$ and $\sigma _{8}$ tensions, suggesting that the
interaction terms might be key to solving some persistent problems in
cosmology.

Furthermore, the models analysed show varied asymptotic behaviours,
indicating the complexity and sensitivity of cosmic evolution to the
specific characteristics of dark sector interactions. We conclude that
understanding these interactions not only provides insights into the nature
of dark matter and dark energy but also helps in predicting the future
dynamics of the universe.

The analysis and results underscore the importance of considering different
interaction models and their implications on cosmological scales. In future
work, we plan to investigate the extent to which such interacting models can
help reduce cosmological tensions. This research lays a foundational
framework for such investigations, highlighting the critical role of
theoretical models in cosmology.

\textbf{Data Availability Statements:} Data sharing is not applicable to
this article as no datasets were generated or analyzed during the current
study.

\begin{acknowledgments}
AP thanks the support of VRIDT through Resoluci\'{o}n VRIDT No. 096/2022 and
Resoluci\'{o}n VRIDT No. 098/2022. KD was funded by the National Research
Foundation of South Africa, Grant number 131604. AA acknowledges that this
work is based on research supported in part by the National Research
Foundation of South Africa (Grant Numbers 151059). Part of this work was
supported by Proyecto Fondecyt Regular 2024, Folio 1240514, Etapa 2024
\end{acknowledgments}

\end{document}